\documentclass[aps,pre,reprint,10pt,twocolumn,nofootinbib,showpacs]{revtex4-1}
\usepackage{graphicx}
\usepackage{amssymb}
\usepackage{amsmath}
\usepackage{amsfonts}
\usepackage{color}
\usepackage[latin1]{inputenc}
\usepackage{float}

\newcommand{\expct}[1]{\langle{#1}\rangle}

\begin{document}
\title{Geometry dependence in linear interface growth}

\author{I. S. S. Carrasco$^{(a,b)}$}
\email{ismael.carrasco@ufv.br}
\author{T. J. Oliveira$^{(a)}$}
\email{tiago@ufv.br}
\affiliation{$(a)$ Departamento de F\'isica, Universidade Federal de Vi\c cosa, 36570-900, Vi\c cosa, Minas Gerais, Brazil \\
$(b)$ Instituto de F\' isica, Universidade Federal Fluminense, Avenida Litor\^ anea s/n, 24210-340 Niter\' oi, Rio de Janeiro, Brazil}

\begin{abstract}
The effect of geometry in the statistics of \textit{nonlinear} universality classes for interface growth has been widely investigated in recent years and it is well known to yield a split of them into subclasses. In this work, we investigate this for the \textit{linear} classes of Edwards-Wilkinson (EW) and of Mullins-Herring (MH) in one- and two-dimensions. From comparison of analytical results with extensive numerical simulations of several discrete models belonging to these classes, as well as  numerical integrations of the growth equations on substrates of fixed size (flat geometry) or expanding linearly in time (radial geometry), we verify that the height distributions (HDs), the spatial and the temporal covariances are universal, but geometry-dependent. In fact, the HDs are always Gaussian and, when defined in terms of the so-called ``KPZ ansatz'' $[h \simeq v_{\infty} t + (\Gamma t)^{\beta} \chi]$, their probability density functions $P(\chi)$ have mean null, so that all their cumulants are null, except by their variances, which assume different values in the flat and radial cases. The shape of the (rescaled) covariance curves is analyzed in detail and compared with some existing analytical results for them. Overall, these results demonstrate that the splitting of such university classes is quite general, being not restricted to the nonlinear ones.
\end{abstract}


\maketitle

\section{Introduction}
\label{secintro}

Interface growth is ubiquitous in nature. The complex morphologies formed in this far from equilibrium process have scales ranging from nano- (e.g., in thin film deposition \cite{ohring,krugbook}) to dozen of meters (e.g. in forest fire fronts), passing through intermediate scales in growth of cell colonies \cite{Huergo2010}, cracks \cite{Maloy2001}, ice deposits \cite{Lowe2007}, paper combustion \cite{Zhang1992}, etc. Despite the very different scales and underlying microscopic processes determining the evolution of such interfaces, they all are expected to present dynamic scaling properties - with their squared width $w_2$ increasing in time as $w_2 \sim t^{2\beta}$, while their correlation length $\xi$ follows $\xi \sim t^{1/z}$ - which allow us to group them in a few number of universality classes (UCs). The most important of these ones being the Kardar-Parisi-Zhang (KPZ) \cite{KPZ}, Villain-Lai-Das Sarma (VLDS) \cite{Villain,LDS}, Edwards-Wilkinson (EW) \cite{EW} and Mullins-Herring (MH) \cite{Mullins1957,Herring1950} UCs (see, e.g, Refs. \cite{barabasi,KrugAdv} for details).

The great majority of studies of growing interfaces in the past decades have focused on calculating the scaling exponents ($\beta$, $z$ and others), in order to determine their UC \cite{barabasi,KrugAdv}. In the last years, however, the attention has been turned to more fundamental quantities such as the (1-point) height distributions (HDs) and (2-point) correlators, which have revealed an interesting dependence of the UCs on the geometry of the system. More specifically, the height $h$ at a given point of a growing interface is expected to evolve according to the ``KPZ ansatz'' \cite{Krug1992,Prahofer2000}
\begin{equation}
 h \simeq v_{\infty} t + (\Theta t)^{\beta} \chi,
 \label{Eqansatz}
\end{equation}
where $v_{\infty}$ is the asymptotic growth velocity of the interface, $\Theta$ sets the interface width amplitude and $\chi$ is a random variable which fluctuates according to universal HDs [$P(\chi)$]. The two-point correlation functions can be defined in general as
\begin{equation}
 C(r,t,t_0) = \left\langle \tilde{h}(\vec{x}+\vec{r},t)\tilde{h}(\vec{x},t_0) \right\rangle,
\end{equation}
where $\tilde{h}\equiv h - \left\langle h\right\rangle $. For equal times, one has the spatial covariance $C_S(r,t) \equiv C(r,t,t_0=t) \simeq w_2 F(r/\xi)$, while for a single point one has the temporal covariance $C_T(t,t_0) \equiv C(r=0,t,t_0) \simeq \sqrt{w_2(t)w_2(t_0)} \mathcal{A}(t/t_0)$, where $F(x)$ and $\mathcal{A}(y)$ are expected to be universal scaling functions. A number of recent works have revealed that $P(\chi)$, as well as $F(x)$ and $\mathcal{A}(y)$ assume different forms for flat and curved geometries. For instance, such geometry-dependence was initially observed in the solution of a one-dimensional (1D) model \cite{Prahofer2000} belonging to the \textit{nonlinear} KPZ UC and, since then, it has been widely confirmed theoretically \cite{Sasamoto2010,Amir,Calabrese2011}, numerically \cite{Alves11,tiago12a,Alves13,HealyCross,silvia17} and experimentally \cite{Takeuchi2010,Takeuchi2011} for other 1D KPZ systems. Generalizations of this scenario for 1D circular KPZ interfaces ingrowing \cite{Fukai2017,Ismael18} or evolving out of the plane \cite{Ismael19} have been also investigated more recently. Furthermore, a similar splitting of $P(\chi)$, $F(x)$ and $\mathcal{A}(y)$ have been numerically observed in the 2D KPZ class \cite{healy12,tiago13,healy13,Ismael14}, as well as in the \textit{nonlinear} VLDS UC in both 1D and 2D \cite{Ismael16a}. Thus, nowadays it is well established that nonlinear UCs for interface growth split into subclasses depending on whether the interfaces are flat or curved.

For the \textit{linear} EW and MH UCs, notwithstanding, a systematic study of such splitting is lacking in the literature. For instance, the amplitudes of $w_2$ - where a difference in HDs for these UCs might arise - have been exactly calculated from the solution of the EW and MH equations [Eqs. \ref{eqEW} and \ref{eqMH} defined below] in the flat case in 1D (see, e.g, Ref. \cite{KrugAdv}) and in 1D circular geometry [Eq. \ref{eqLinRadial} defined below] for the EW class \cite{Singha2005,Masoudi1}. However, as far as we know, for the 1D circular MH class, as well as for both classes and geometries in 2D they have never been reported elsewhere. For the spatial covariances, approximated analytical results exists for their behavior in the large $r$ limit for the flat geometry in both 1D and 2D \cite{Majaniemi}, but for the radial cases it seems that neither analytical predictions nor numerical analysis exist for $F(x)$. For the temporal covariances, there are analytical calculations for their general behavior in the flat case \cite{KrugKallabis97}, as well as for 1D circular interfaces \cite{Singha2005}, whereas in the 2D radial case only the form of the asymptotic decay [$\mathcal{A}(y) \sim y^{-\bar{\lambda}}$] is known analytically \cite{Singha2005}. However, numerical confirmations of the universality of such behaviors, as well as of most of the other analytical results mentioned above seem to be absent in the literature.

Hence, the aim of the present work is gathering the existing results for the statistics of linear interfaces together, demonstrate their universality and fill the several existing gaps. Through this analysis, a throughout confirmation of the existence of a geometry dependence in the linear UCs is given. In order to do so, we investigate both analytically and numerically several models belonging to the EW and MH classes, on 1D and 2D substrates, in the flat and radial geometries. 

The rest of the paper is organized as follows. In Sec. \ref{secModels} we present the models and numerical methods employed to investigate them. In Secs. \ref{secHDs}, \ref{secCovS} and \ref{secCovT} results for HDs, spatial and temporal covariances are reported, respectively. Our conclusions and final discussions are summarized in Sec. \ref{secConc}. In appendix \ref{apExSol} the analytical solution of the radial linear equations is devised. A discussion on the inverse method is provided in appendix \ref{apInvMet}.

\section{Models and Numerical Methods}
\label{secModels}

Let us recall that in the continuum limit the dynamics of interfaces belonging to EW and MH classes are described by the EW equation
\begin{equation}
\frac{\partial h(\vec{x},t)}{\partial t} = F + \nu_2 \nabla^2 h + \eta(\vec{x},t)
\label{eqEW}
\end{equation}
and the MH equation
\begin{equation}
\frac{\partial h(\vec{x},t)}{\partial t} = F - \nu_4 \nabla^4 h + \eta(\vec{x},t),
\label{eqMH}
\end{equation}
respectively, where $F$ can be viewed as the net particle flux per site, $\nu_2$ and $\nu_4$ are phenomenological parameters, and $\eta$ is a Gaussian white noise, with zero mean and correlation $\expct{\eta(x,t)\eta(x',t')} = 2 D \delta^{d_s}(x-x')\delta(t-t')$, being $d_s$ the substrate dimension. In order to investigate them in radial geometry, these equations must be rewritten in appropriate coordinates [see Eq. \ref{eqLinRadial} in Appendix \ref{apExSol} for the 1D case]. We can numerically integrate Eqs. \ref{eqEW} and \ref{eqMH} by defining them on discretized substrates, which will be an array of $L$ sites in 1D and a rectangular lattice with $L_x \times L_y$ sites in 2D. Thenceforth, we make the height at a given site ${\vec{x}}$ to evolve, following the Euler method, as \cite{Moser1991}
\begin{equation}
h_{\vec{x}}(t+\delta t)=h_{\vec{x}}(t)+ (\delta t) \nu_z \mathcal{L}[h_{\vec{x}}(t)]+\sqrt{24 (\delta t) D}\,\mathcal{R}\;,
\end{equation}
where $z=2$ (EW) or $4$ (MH), and we have made $F=0$, without any loss of generality. The functions $\mathcal{L}[h_{\vec{x}}(t)]$ are given, in 2D, by
\begin{align}\nonumber
\begin{split}
&\mathcal{L}_{EW} \equiv \frac{h_{i+1,j} + h_{i-1,j} + h_{i,j+1} + h_{i,j-1} - 4 h_{i,j}}{(\Delta x)^2 (\Delta y)^2},\\
&\mathcal{L}_{MH} \equiv -[h_{i+2,j} + h_{i-2,j} - 4h_{i+1,j} - 4h_{i-1,j} + h_{i,j+2} \\
&+ h_{i,j-2} - 4h_{i,j+1} - 4h_{i,j-1} + 12 h_{i,j}]/[(\Delta x)^4 (\Delta y)^4]\; ,
\end{split}
\end{align}
which are discretized approximations for $\nabla^2h_{i,j}$ and $\nabla^4h_{i,j}$, respectively. The simplification of these expressions for 1D is straightforward. $\Delta x$ and $\Delta y$ can be set to $1$, without any loss of generality. $\mathcal{R}$ is a random variable sorted from a uniform distribution in the interval $(-1/2, 1/2)$. Moreover, we use $\delta t=0.001$ in 1D and $\delta t=0.01$ in 2D, and two set of parameters ($\nu_z,D$), which are displayed in Tab. \ref{tab1}. For sake of simplicity, hereafter, we will refer to them as EW$_I$, EW$_{II}$, MH$_I$ and MH$_{II}$ models.

\begin{table}
	\caption{Parameters used in the numerical integration of the EW and MH equations, in both 1D and 2D. We have set $F=0$ in all cases.}
	\begin{center}
		\begin{tabular}{c c c c c c c c c}
			\hline \hline
			                & & EW$_I$  & &  EW$_{II}$  & &  MH$_I$ & &  MH$_{II}$ \\
			\hline
			1D - $\nu_z$    & &    1    & &      6      & &     1   & &    6       \\
			1D - $D$        & &    0.25    & &      1      & &     0.25   & &    1       \\
			\hline
			2D - $\nu_z$    & &    1    & &      2.5      & &     1   & &    2.5       \\
			2D - $D$        & &    1    & &      1      & &     1   & &    1       \\
			\hline\hline
		\end{tabular}
		\label{tab1}
	\end{center}
 
\end{table}

The discrete models investigated in the EW class are the symmetric single step (SSS) \cite{Meakin1986} and the Family \cite{Family} models, while in the MH class two versions of the large curvature (LC) model \cite{LCM,KrugLCM} are analyzed. In all models, deposition (and evaporation in the SSS model) occurs sequentially at randomly chosen sites. The growth rules at a given site, say $i$, are as follows:
\begin{itemize}
\item \textit{SSS}: $h_i\rightarrow h_i + 2$ if $\Delta h \equiv (h_j - h_i)=1$ $\forall$ nearest neighbors (NN's) $j$; or $h_i\rightarrow h_i - 2$ if $\Delta h=-1$  $\forall$ NN's $j$; or the deposition/evaporation attempt is rejected.

\item \textit{Family}: $h_i\rightarrow h_i + 1$ if $h_i \leq h_j$  $\forall$ NN's $j$. Otherwise, the NN $j$ with minimal height is taken and $h_{j}\rightarrow h_{j} + 1$. A random draw resolves possible ties.

\item \textit{LC1}\cite{LCM}: The local curvature $C_k$ [e.g, $C_k \equiv h_{k+1} + h_{k-1}-2h_k$ in 1D] is calculated for $k=i$ and $k=j$ $\forall$ NN's $j$ and then $h_i\rightarrow h_i + 1$ if $C_i \geq C_{j}$ $\forall$ NN's $j$. Otherwise, the NN $j$ with maximal curvature is taken and $h_{j}\rightarrow h_{j} + 1$. A random draw resolves possible ties.

\item \textit{LC2}\cite{KrugLCM}: The growth rule is identical to LC1, but instead of sorting a random site $i$, supposing that the sites live between the lattice vertices, we sort a random vertex. Therefore, in 1D we have to compare $C_k$ only for two sites ($k=i$ and $k=i+1$). Similarly, in 2D we look at the four sites around a given vertex.

\end{itemize}

To enable uniform aggregation and evaporation in the SSS model, we start the growth with a checkerboard initial condition, so that sites with height $h(t=0)=-1/2$ are surrounded by NN sites with $h(t=0)=1/2$ and vice-versa. For the other models one simply makes $h_i(t=0)=0$ $\forall$ $i$. The time is defined so that we attempt to deposit one monolayer in a time unity. Hence, for a substrate of fixed size $L_0$ one has $\Delta t = 1/L_0$. Since all deposition attempts are accepted in the Family and LC models, their interfaces have mean heights increasing deterministically as $\left\langle h\right\rangle = t$, so that $v_{\infty}=1$ (in Eq. \ref{Eqansatz}) for these models. On the other hand, in the SSS model and integrations of the linear equations, the mean height of a given interface stochastically fluctuates around zero, without any tendency to increase or decrease in time, leading to $\left\langle h\right\rangle = 0$ and, so, $v_{\infty} = 0$. For all models, one may conclude also that $\left\langle \chi \right\rangle = 0$ (in Eq. \ref{Eqansatz}).

In the form just described, the models and integrations of the growth equations yield asymptotically flat interfaces and, so, the \textit{flat} geometry. As demonstrated in a number of recent works on nonlinear interface growth \cite{Ismael14,HHTake2015,Ismael16a,Ismael18} and confirmed in the following for linear ones, a simple way to investigate these systems in the \textit{radial} geometry is by considering their growth on substrates whose size enlarges stochastically in time, in each direction, as $L(t) = L_0 +\omega t$, where $L_0$ is the initial size and $\omega$ is the enlargement rate . To do so, we randomly mix the growth rules defined above [occurring with probability $P_g(t)=N(t)/(N(t)+\omega d_s)$] with the substrate enlargement [with probability $P_{\omega}(t)=\omega d_s/(N(t)+\omega d_s)$]. Here, $N = L$ in 1D and $N=L_x L_y$ in 2D. The enlargement is implemented by randomly choosing a column (say $i$) and duplicating it. Namely, a new column with height $h_i$ is created between $i$ and the older $i+1$ one. Each event increases the time by $\Delta t=1/(N+\omega d_s)$. For the SSS model, one has to duplicate a pair of neighboring columns in order to not violate the single step restriction $|\Delta h|=1$ and, then, $\omega$ is changed to $\omega/2$ in the expressions above. Note that in 2D we have $\left\langle L_x \right\rangle = \left\langle L_y \right\rangle$, so that on average one has square lattices expanding linearly in time. We remark that with such definitions for the probabilities and $\Delta t$, one still has $\left\langle h \right\rangle = t$ for the Family and LC models, and $\left\langle h \right\rangle = 0$ for the other ones. Thence, independently of the geometry, the models investigated here have $\left\langle \chi \right\rangle = 0$ and $v_{\infty} = 1$ (Family and LC) or $v_{\infty} = 0$ (the other models).

As we have demonstrated for KPZ systems \cite{Ismael14,Ismael18,Ismael19}, large $L_0$'s introduce crossover effects in expanding systems, from the flat to the truly asymptotic radial statistics and this may hamper the analysis of last one. So, in order to avoid undesired transients, here we will use $L_0=4$ in all simulations for enlarging substrates. Moreover, it is well-known from previous studies that no quantity is affected in the asymptotic regime by the rate $\omega$, at least when the substrate expansion is linear in time \cite{Ismael19}, which is the case here. In fact, we have confirmed that all results presented in the following are indeed $\omega$-independent and, then, only data for $\omega=12$ and $\omega=2$ will be shown for 1D and 2D, respectively. For fixed substrate sizes ($\omega=0$), we will consider $L=L_0$ up to $2^{17}$ for the discrete models and $2^{13}$ for the integrations in 1D; and $L$ up to $2^{10}$ for all cases in 2D. At least $\sim 10^7$ substrate sites are considered in the statistics for each system in a given time.

\section{Height fluctuations in linear growth}
\label{secHDs}

In this section we focus on the behavior of the asymptotic height distributions (HDs), during the growth regime, analyzed in the light of the KPZ ansatz (Eq. \ref{Eqansatz}). Let us remark that for the EW ($z=2$) and MH ($z=4$) classes the squared interface width behaves as $w_2 \simeq \left( D/\nu_z^{1-2\beta}\right) b t^{2\beta}$ \cite{KrugAdv}, where $b$ is expected to be a universal constant. Meanwhile, from Eq. \ref{Eqansatz} one has $w_2 \simeq \Theta^{2\beta} \left\langle \chi^2\right\rangle_c  t^{2\beta}$, so that one may identify $b = \expct{\chi^2}_c$ and
\begin{equation}
 \Theta = \left( \frac{D}{\nu_z^{1-2\beta}}\right)^{\frac{1}{2\beta}},
\label{Eqtheta}
\end{equation}
where $2\beta=1-d_s/z$, whilst the dynamic exponent is always $z=2$ for EW and $z=4$ for MH class, regardless the substrate dimension $d_s$ \cite{KrugAdv,barabasi}.
Since the EW (Eq. \ref{eqEW}) and MH (Eq. \ref{eqMH}) equations are linear, the HDs in these classes are Gaussian, in both 1D and 2D. Thereby, the single non-null cumulant of $P(\chi)$ is its variance $\left\langle \chi^2\right\rangle_c$, once as noted above $\left\langle \chi \right\rangle = 0$, and $\left\langle \chi^n\right\rangle_c = 0$ for $n>2$ for Gaussian distributions. So, any difference between HDs for flat and radial geometries might appear in $\left\langle \chi^2\right\rangle_c$. This contrasts with the nonlinear classes, where the full probability density functions $P(\chi)$ are different.

\subsection{Results for 1D interfaces}

The exact solution of the 1D EW equation on a flat substrate yields $w_2 = \left( \frac{D}{\sqrt{\nu_2}} \right) \left( \frac{\sqrt{2}\Gamma(1/2)}{\pi}\right) t^{1/2}$ (see, e.g., Ref. \cite{KrugAdv}\footnote{Note that the noise amplitude $D$ in \cite{KrugAdv} is defined as $2D$ here.}), where $\Gamma(x)$ is the gamma function. Therefore, one may identify $\beta=1/4$, $\Theta=D^2/\nu_2$ (in agreement with Eq. \ref{Eqtheta}) and $\left\langle \chi^2\right\rangle_{c,EW,1D}^f = \left( \frac{\sqrt{2} \Gamma(1/2)}{\pi}\right) \simeq 0.79788$. On the other hand, the solution of the radial 1D EW equation gives $w_2 = \left( \frac{D}{\sqrt{\nu_2}} \right) \sqrt{\frac{\pi}{2}} t^{1/2}$ \cite{Singha2005,Masoudi1}, showing that the variance of the HDs changes to $\left\langle \chi^2\right\rangle_{c,EW,1D}^r = \sqrt{\pi/2} \simeq 1.25331$, while the other quantities are still the same. For the 1D MH equation on flat substrates, it is known that $w_2 = \left( \frac{D}{\nu_4^{1/4}} \right) \left( \frac{2^{3/4}\Gamma(1/4)}{3\pi}\right) t^{3/4}$ \cite{KrugAdv}, leading to $\beta=3/8$, $\Theta=(D^4/\nu_4)^{1/3}$ (which is again consistent with Eq. \ref{Eqtheta}) and $\left\langle \chi^2\right\rangle_{c,MH,1D}^f = \left( \frac{2^{3/4}\Gamma(1/4)}{3\pi}\right) \simeq 0.64697$. For the radial 1D MH equation, notwithstanding, although its solution has been discussed in some works \cite{EscuderoMH1,EscuderoMH2}, it seems that the amplitude of $w_2$ has not been reported elsewhere. So, we solve this equation here, see the appendix \ref{apExSol}, and find that $\expct{\chi^2}_{c,MH,1D}^r = \pi/[54^{1/4}\Gamma(3/4)] \approx 0.94573$, which confirms the geometry-dependence in the HDs' variance also in this class.

\begin{figure}[!t]
	\includegraphics[width=4.25cm]{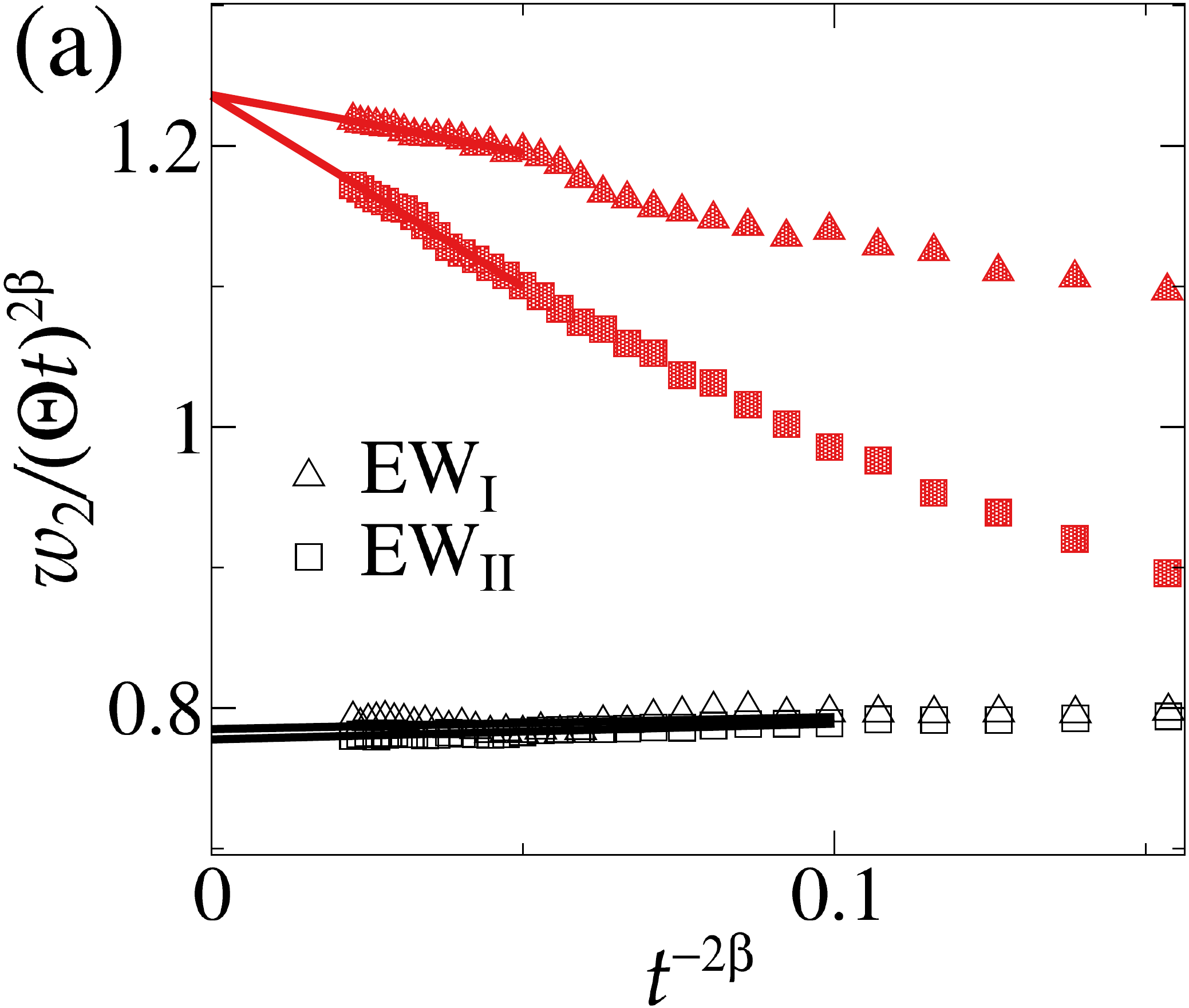}
	\includegraphics[width=4.25cm]{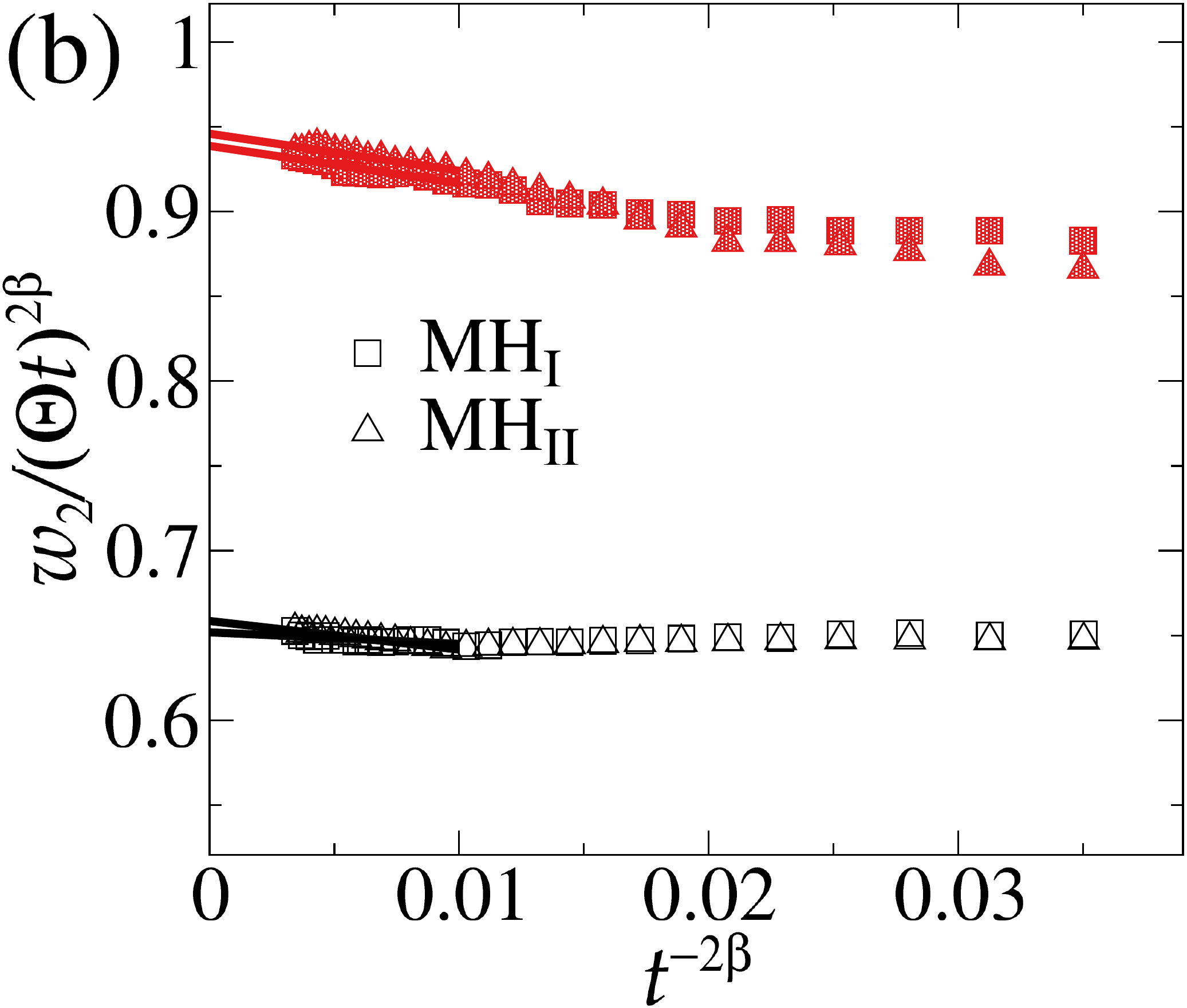}
	\includegraphics[width=4.25cm]{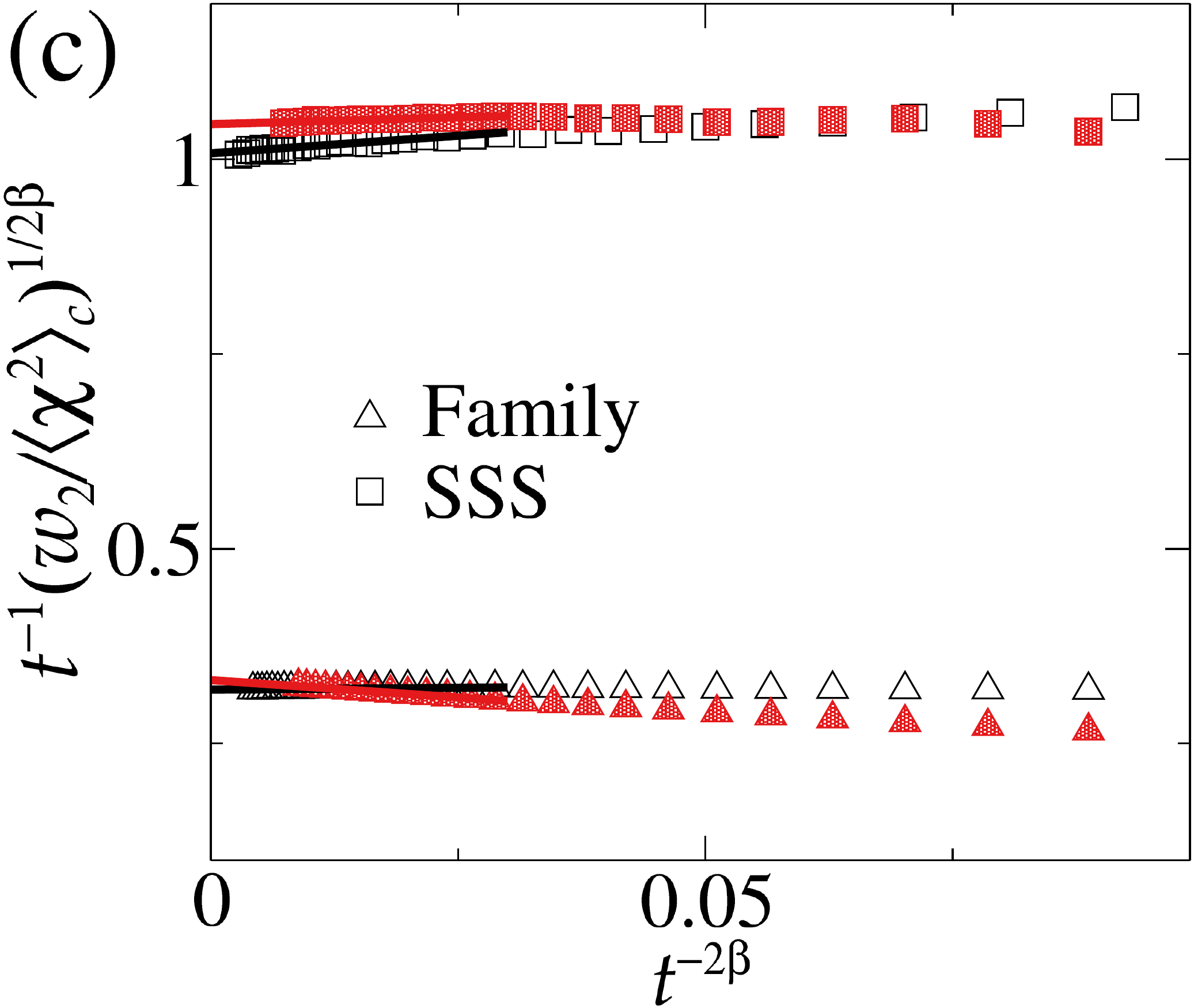}
	\includegraphics[width=4.25cm]{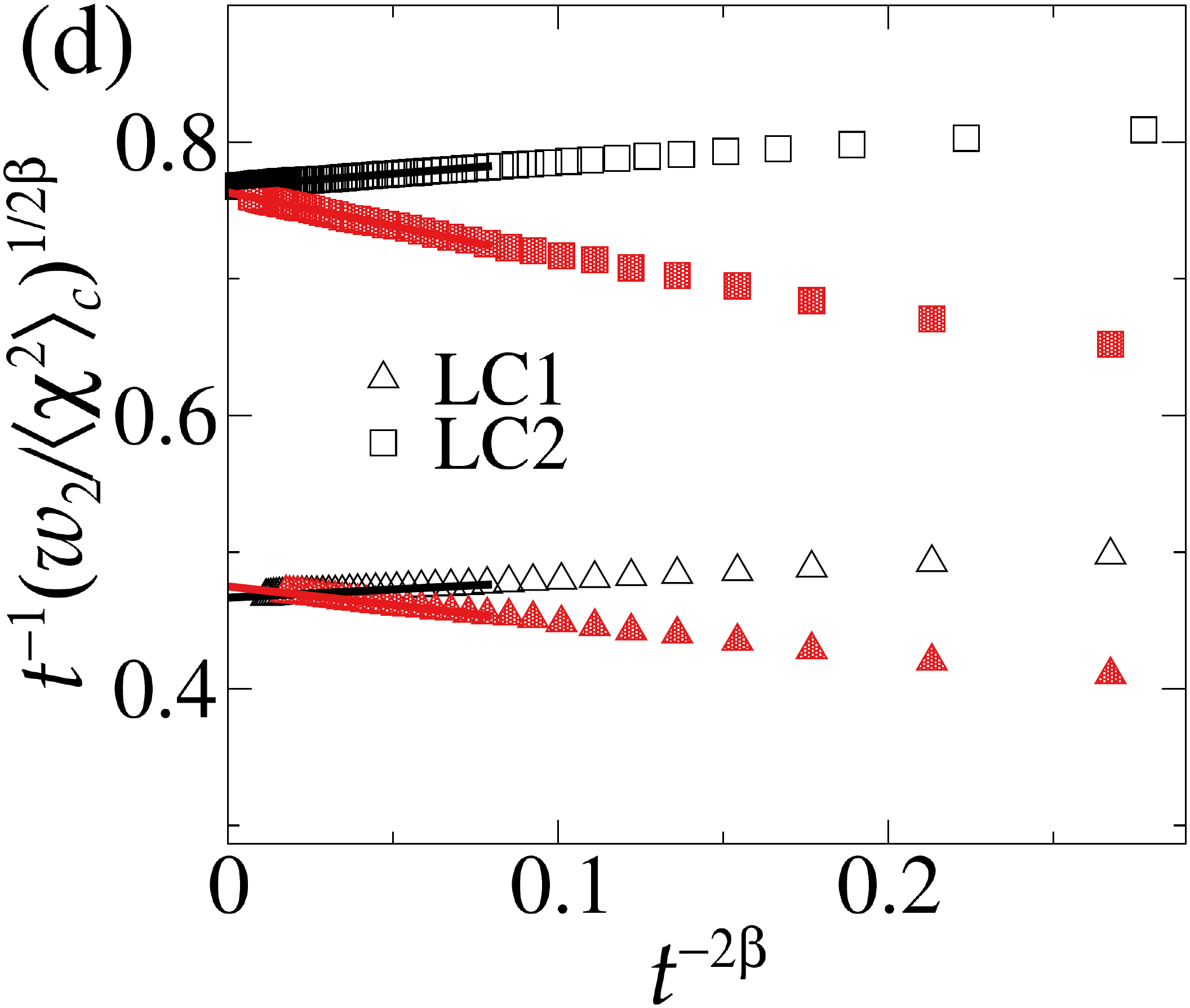}
	\caption{Extrapolations to the long time limit of the rescaled squared interface widths for the different models and classes in 1D: $w_2/(\Theta t)^{2\beta}$ as function of $t^{-2\beta}$ for the integrations of the 1D EW (a) and MH (b) equations, for two set of parameters each; and $[w_2/\langle \chi^2\rangle_c]^{1/2\beta}/t$ versus $t^{-2\beta}$ for the Family and SSS models (c), and the LC models (d). In all cases, data for fixed-size (open - black) and expanding (full symbols - red) substrates are shown. The solid lines represent the linear fits used to perform the extrapolations.}
	\label{fig1}
\end{figure}

In order to verify that our integration method yields reliable results and, more important, that simulations on substrates enlarging linearly in time are indeed consistent with the radial geometry, we numerically integrate the 1D growth equations (\ref{eqEW} and \ref{eqMH}). Figures \ref{fig1}(a) and \ref{fig1}(b) show the extrapolation of $w_2/(\Theta t)^{2\beta}$ for the long-time limit, for the EW and MH classes, respectively, where results for both fixed-size and enlarging substrates are shown. According to Eq. \ref{Eqansatz}, the extrapolated values are expected to converge to the values of $\expct{\chi^2}_c$ reported above for each class and geometry. In fact, in all cases the relative difference between the central values from extrapolations and the exact ones does not surpass 2\%, and they always agree within the error bars.

Thanks to a mapping of the 1D SSS model on the kinetic Ising model, one knows that it is described in the continuum limit by the EW equation with $\nu_2=D=1$ \cite{Plischke} and, so, $\Theta=D^2/\nu_2=1$. Moreover, by construction, for the Family and LC models one has $D=1/2$ \cite{KrugAdv}, whereas the parameters $\nu_z$, with $z=2,4$, are not exactly known for these models and the scarce numerical estimates of them are not so accurate. For instance, the value $\nu_4 \approx 0.14$ was found for the LC2 model by comparing numerical results (for the flat case) with the exact result above for $w_2$ and the one for the average local slope $\expct{(\nabla h)^2} \equiv \expct{(h_{i+1}-h_i)^2}$ \cite{KrugAdv}. Moreover, $\nu_2 \sim 0.8$ was obtained for the Family model with a pseudospectral inverse method \cite{Giacometti}. For the best of our knowledge, no estimate exists for the LC1 model. We remark that, when applying inverse methods \cite{Lam,Giacometti}, one usually supposes that a given model (or a real surface) is described by a given growth equation and try to find all its coefficients, e.g, $F$, $\nu_z$ and $D$ in EW and MH equations. See appendix \ref{apInvMet} for details. Notwithstanding, since one knows that $F=1$ and $D=1/2$ for the Family and LC models, we can apply the inverse method more effectively by letting such parameters fixed and calculating only $\nu_z$. By doing this, we have indeed found more accurate values, which are summarized in Tab. \ref{tab2}.

\begin{table}[!t]
	\caption{\textit{Top:} Coefficients $\nu_z$ from the inverse method, with exception of the exact value for the SSS model, and the exact values for $D$ for the discrete EW (left) and MH (right) models in 1D. The values of $\Theta$ obtained using Eq. \ref{Eqtheta} are also shown. \textit{Bottom:} Extrapolated values of $\Theta$ from Figs. \ref{fig1}(c) and \ref{fig1}(d), for flat and radial systems.}
	\begin{center}
		\begin{tabular}{c c c c c c c|c c c c c }
			\hline \hline
			                                    & &  Family   & & SSS        & &    & &  LC1      & &  LC2       &    \\
			\hline
			$\nu_z$                             & &  0.78(2)  & &  1         & &    & &  0.61(2)   & &  0.142(6)   &    \\
			$D$                                 & &  1/2      & &  1         & &    & &  1/2       & &  1/2        &    \\
			$\Theta$                            & &  0.321(8) & &  1         & &    & &  0.468(5)  & &  0.761(9)   &    \\
			\hline
			$\Theta$ (flat)                     & &  0.315(1) & &  1.003(4)  & &    & &  0.471(1)  & &  0.765(1)   &    \\
			$\Theta$ (radial)                   & &  0.318(2) & &  1.01(1)   & &    & &  0.468(2)  & &  0.763(2)   &    \\
			\hline\hline
		\end{tabular}
		\label{tab2}
	\end{center}
\end{table}

Now, we may verify the universality of the HDs' variances, e.g, by calculating $\Theta$ from the values of $\nu_z$ and $D$ for the discrete models (see Tab. \ref{tab2}) and comparing them with the values coming from extrapolations of $\left[ w_2/\expct{\chi^2}_c\right]^{1/2\beta}/t$ for $t\rightarrow \infty$, using the exact values of $\expct{\chi^2}_c$. This is done in Figs. \ref{fig1}(c) and \ref{fig1}(d) for all discrete models. The values of $\Theta$ estimated in this way are also displayed in Tab. \ref{tab2}, which agree with the expected ones (obtained from $\nu_z$ and $D$) within the error bars, giving a compelling demonstration of the universality of the HDs for 1D linear interfaces. It is worth remarking that a given model is expected to have the same $\Theta$ (actually, the same $\nu_z$ and $D$) regardless the substrate size is fixed or enlarging linearly in time \cite{Ismael14}. An exception to this occurs for circular interfaces evolving out of the plane, when $L$ is a nonlinear function of $t$ \cite{Ismael19}, which is not the case here. Therefore, the agreement of the $\Theta$'s for flat and radial systems observed in Tab. \ref{tab2} is per se a confirmation of the correctness (i.e, universality) of the values of $\expct{\chi^2}_c$ used in the extrapolations. This fact will be very important in what follows in the analysis of 2D interfaces.

\subsection{Results for 2D interfaces}

Although in 2D and higher dimensions we can deal with the linear equations to obtain the exact scaling exponents, as far as we know, it is not possible to perform the summations appearing along the solutions (see appendix \ref{apExSol}) to calculate the exact values of $w_2$'s amplitudes. In fact, it seems that they have never appeared elsewhere. So, we will rely on numerical calculations of these amplitudes to verify their dependence on the geometry, a task which has also not been tackled so far.

\begin{figure}[!t]
	\includegraphics[width=4.25cm]{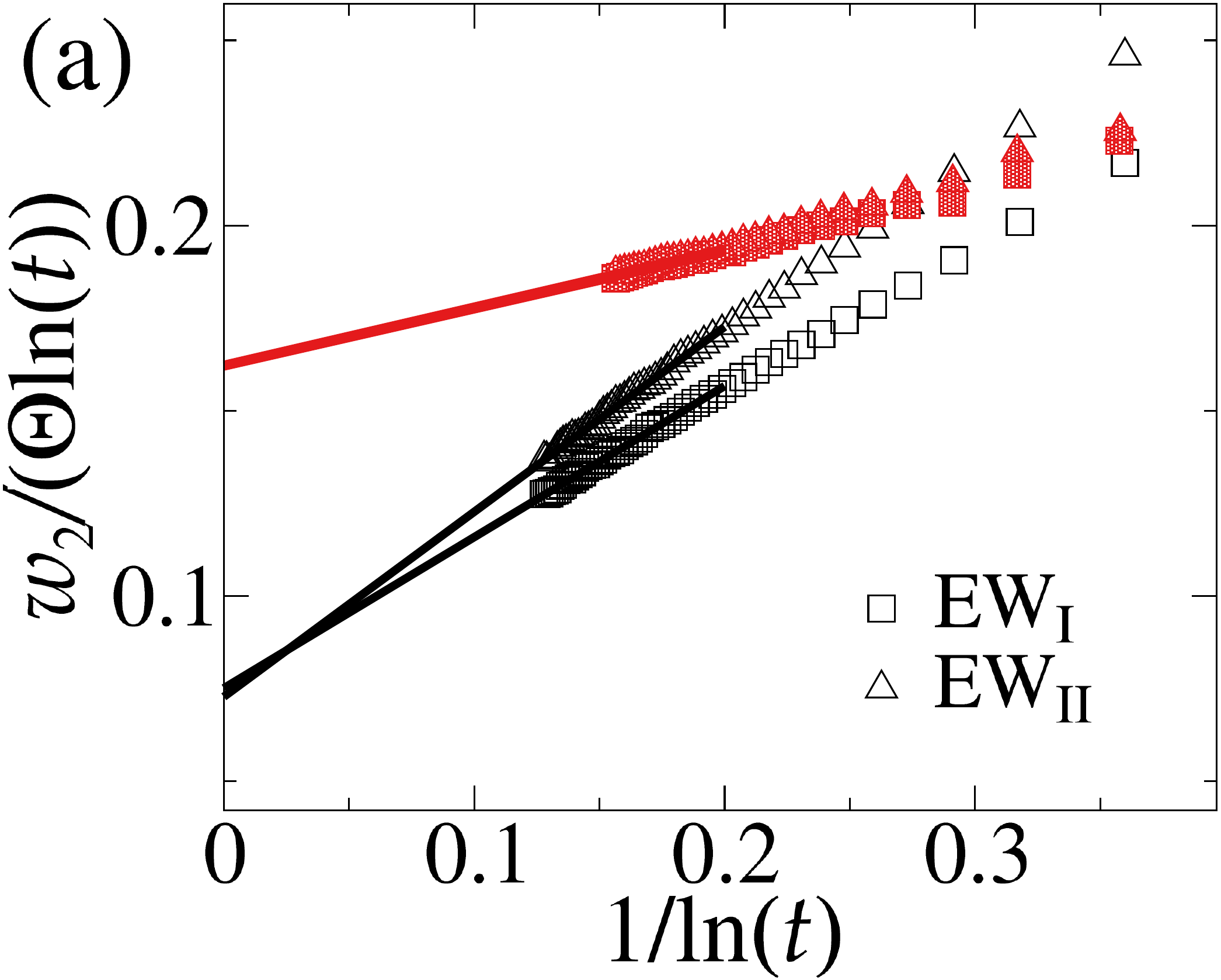}
	\includegraphics[width=4.25cm]{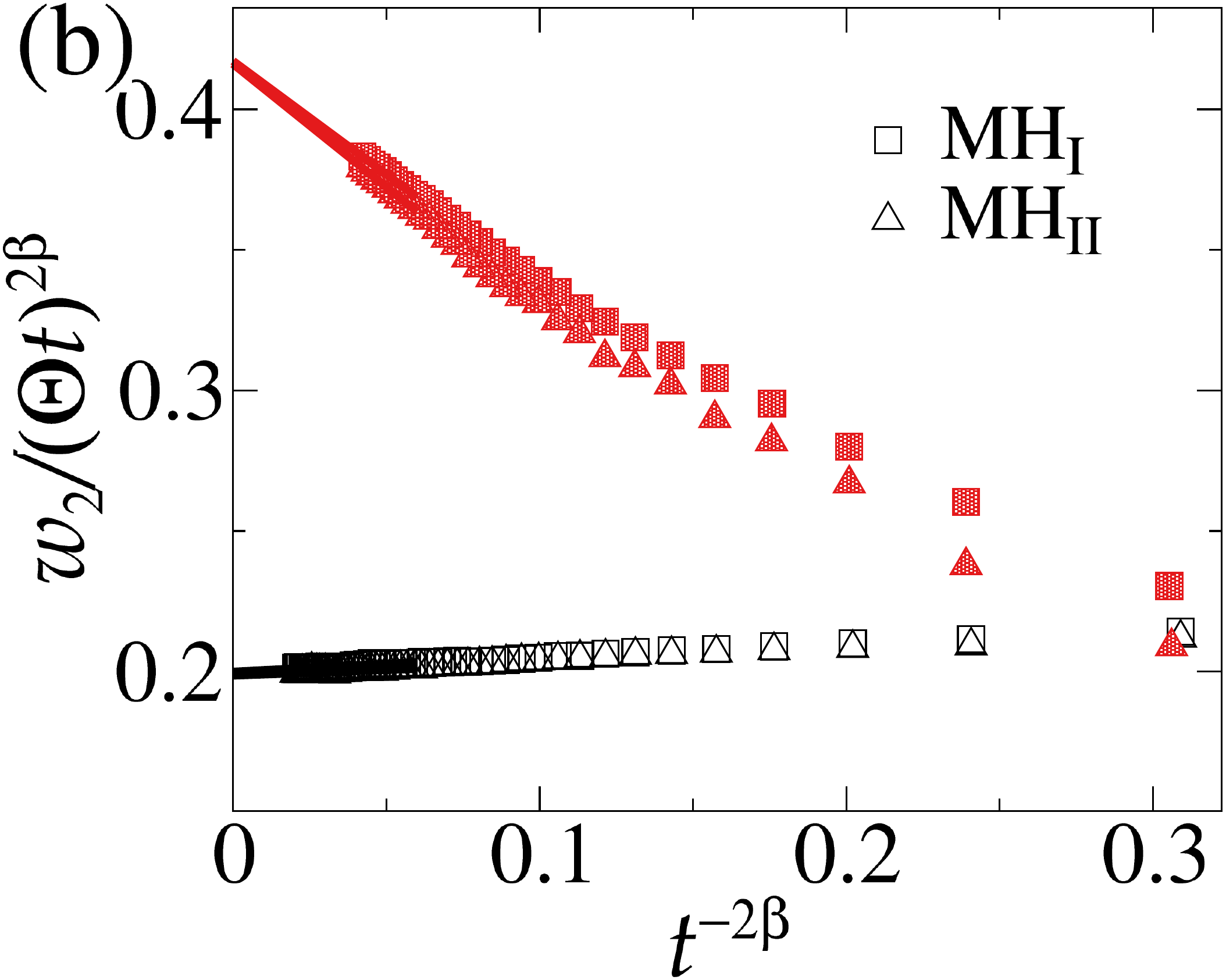}
	\includegraphics[width=4.25cm]{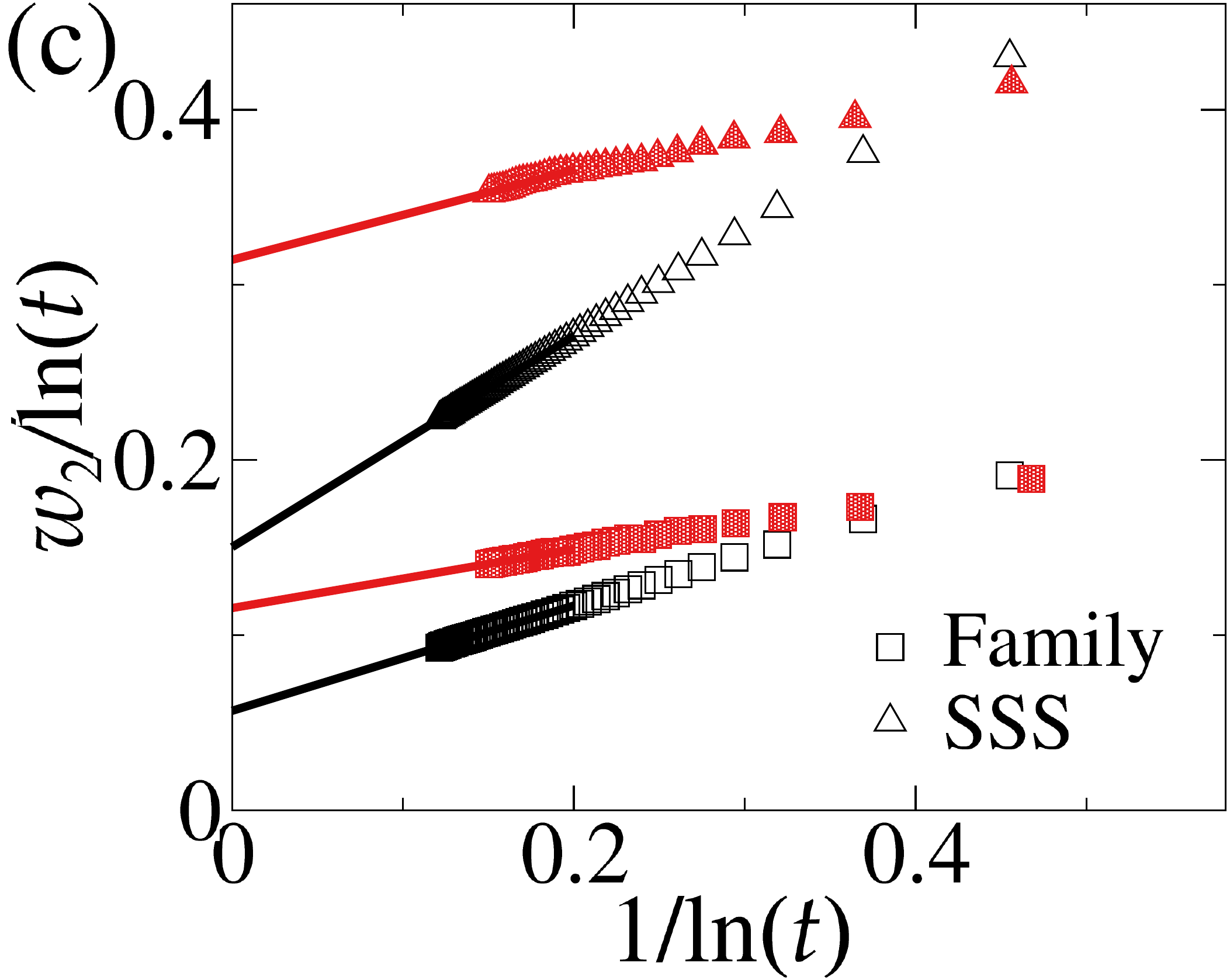}
	\includegraphics[width=4.25cm]{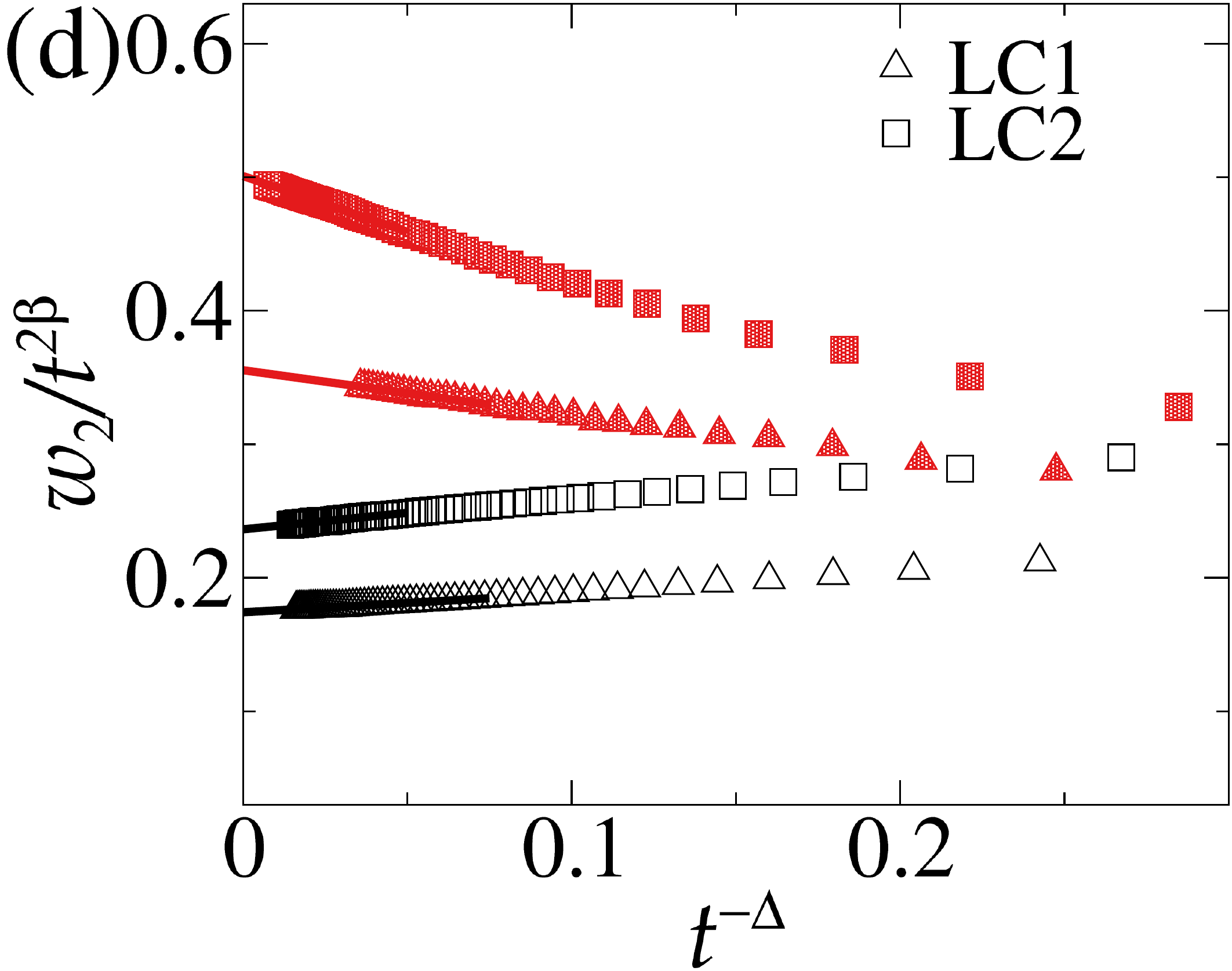}
	\caption{Extrapolations to the long time limit of rescaled squared interface widths for the different models and classes in 2D: (a) $w_2/[\Theta \ln(t)]$ against $1/\ln t$ for the integrations of the 2D EW equation; (b) $w_2/(\Theta t)^{2\beta}$ as function of $t^{-2\beta}$ for the integrations of the 2D MH equation; (c) $w_2/ \ln(t)$ versus $1/\ln t$ for the Family and SSS models; and (d) $w_2/t^{2\beta}$ versus $t^{-\Delta}$ for the LC models. In all cases, data for fixed-size (open - black) and expanding (full symbols - red) substrates are shown. The solid lines represent the linear fits used to perform the extrapolations. In (d) we use $\Delta = 2\beta$ in all cases, with exception of the LC2 model on expanding substrates, where $\Delta = 3\beta$ provides a better linearization.}
	\label{fig2}
\end{figure}

Let us start recalling that $d_s=2$ is the upper critical dimension for the EW class, meaning that $\beta=0$ there, so that the power-law scaling gives place to a logarithmic variation of $w_2$ in time. More specifically, $w_2 \simeq (D/\nu_2)\expct{\chi^2}_c \ln(\nu_2 t/a)$ \cite{KrugAdv}, where $a$ is the lattice parameter ($a=1$ for the models investigated here). Therefore, the ansatz as defined in Eq. \ref{Eqansatz} cannot be applied in this case, once it yields $w_2 \sim t^{2\beta}$. We can however redefine it for the EW class in 2D as
\begin{equation}
 h \simeq v_{\infty} t + \sqrt{\Theta \ln(\nu_2 t/a)} \chi,
\label{EqansatzEW}
\end{equation}
where again $\expct{\chi}=0$ and $\expct{\chi^n}_c=0$ for $n>2$, and $\Theta=D/\nu_2$. Thereby, the HDs' variances can be found by extrapolating $w_2/[\Theta \ln(t)]$ for $t \rightarrow \infty$. Figure \ref{fig2}(a) shows such extrapolations for the integrations of the 2D EW equation on fixed-size and expanding substrates. The values of $\expct{\chi^2}_c$ for each condition are displayed in Tab. \ref{tab3}, which agree quite well for both set of parameters considered in the integrations, but are different for flat and expanding systems. We notice that the extrapolation of $w_2/[\Theta \ln(\nu_2 t)]$, including $\nu_2$, returns essentially the same values from Tab. \ref{tab3}, as expected.

\begin{table}[!t]
	\caption{\textit{Top:} Extrapolated values of HDs' variances $\expct{\chi^2}_c$ for the integration of the growth equations in 2D, from Figs. \ref{fig2}(a) and \ref{fig2}(b). \textit{Bottom:} Extrapolated values of $g_2$ (see definitions in the text) from Figs. \ref{fig2}(c) and \ref{fig2}(d). The corresponding roughness amplitudes $\Theta$ for the discrete models in EW (left) and MH (right) classes in 2D are also shown.}
	\begin{center}
		\begin{tabular}{c c c c c|c c c c }
			\hline \hline
			                            & & $EW_{I}$ & & $EW_{II}$ & & $MH_I$ & & $MH_{II}$\\
			\hline
			$\expct{\chi^2}_c$ (flat)   & &   0.077(2)    & &    0.075(3)    & &   0.200(2)  & &  0.198(2)    \\
			$\expct{\chi^2}_c$ (radial) & &   0.160(1)    & &    0.161(2)    & &   0.415(3) & &  0.416(3)  \\
			\hline\hline
			                            & &  Family  & &    SSS    & &  LC1  & &  LC2\\
			\hline
			$g_2$ (flat)                & &   0.055(1)    & &    0.1497(4)    & &  0.173(1)  & &  0.237(1)\\
			$g_2$ (radial)              & &   0.115(1)    & &    0.322(4)    & &  0.356(2)  & &   0.500(1) \\
			$\Theta$ (flat)   			& &   0.72(4)    & &    1.97(8)    & & 0.76(2)  & &  1.42(4)\\
			$\Theta$ (radial)   		& &   0.71(2)    & &    2.00(5)    & & 0.74(2)  & &   1.45(3)\\
			\hline
			$\nu_z$  (flat)				& &   0.69(4)    & &    0.51(2)    & &   0.176(5) & &  0.33(1)\\
			$\nu_z$ (radial)			& &   0.70(2)   & &    0.50(1)    & &   0.172(3)  & &  0.340(9)\\
			\hline\hline
		\end{tabular}
		\label{tab3}
	\end{center}
\end{table}

For the MH class, the ansatz in Eq. \ref{Eqansatz} is still valid in 2D, so we can proceed similarly to the 1D case, by extrapolating $w_2/(\Theta t)^{2\beta}$, to estimate $\expct{\chi^2}_c$. See Fig. \ref{fig2}(b). The obtained values from the integration of the 2D MH equation are also summarized in Tab. \ref{tab3} and, again, they agree for different parameters $\nu_4$ and $D$ and depend on whether the substrate size is fixed or enlarging. Thus, both EW and MH classes display a geometry-dependence in the HDs' variances also in 2D.

Unfortunately, since the 2D linear interfaces are still very smooth (i.e, $w_2$ is still very small) even at long deposition times, the inverse method fails in returning reliable values for $\nu_z$ and $D$. Hence, to verify the universality of the HDs we extrapolate $g_2 \equiv w_2/\ln t$ (for EW) and $g_2 \equiv w_2/t^{2\beta}$ (for MH class) for $t \rightarrow \infty$, as done in Figs. \ref{fig2}(c) and \ref{fig2}(d), respectively. The extrapolated values for $g_2$ are summarized in Tab. \ref{tab3}. According to the ansatz for 2D EW class (Eq. \ref{EqansatzEW}), $g_2 \rightarrow \Theta \expct{\chi^2}_c$, while from the ``KPZ ansatz'' (Eq. \ref{Eqansatz}), we have $g_2 \rightarrow \Theta^{2\beta} \expct{\chi^2}_c$ for the MH class. Therefore, from the estimates of $g_2$ and $\expct{\chi^2}_c$ in Tab. \ref{tab3}, one may calculate the amplitudes $\Theta$. Such values, also shown in Tab. \ref{tab3}, agree quite well for both flat and radial geometries, for a given model, providing a strong evidence of the universality of the HDs' variances also in 2D. Finally, we notice that it is quite reasonable to expect that the noise amplitudes for the discrete models do not change with dimension. Thence, assuming that $D=1/2$ for the Family and LC models, and $D=1$ for the SSS model, we obtain the values of $\nu_z$ displayed in Tab. \ref{tab3}, whose reliability will be confirmed in the next section.

\section{Spatial covariances}
\label{secCovS}

Now, we discuss the two-point height correlation function, aka spatial covariance $C_S(r,t) = \left\langle \tilde{h}(\vec{x}+\vec{r},t) \tilde{h}(\vec{x},t) \right\rangle \simeq w_2 F(r/\xi(t)) $, focusing on the dependence of the scaling function $F(s)$ with geometry. Note that $C_S(r=0,t) = w_2$, so that $F(0)=1$ and, moreover, for $s\equiv r/\xi \rightarrow 0$ it is expected that $F(s)=1-\mathcal{O}(s^{2\alpha})$, where $\alpha=\beta z$ is the roughness exponent. The behavior of $F(s)$ in the large $s$ limit was approximately calculated for the EW and MH classes for flat geometry in  both 1D and 2D and appears to be given in general by \cite{Majaniemi}
\begin{equation}
 F(s) \sim s^{-\gamma} e^{-c s^{\delta}},
\label{eqScalFuncC_S}
\end{equation}
where the exponents $\delta=z/(z-1)$ and $\gamma=1+\delta/2$ were derived for 1D \cite{Majaniemi}. In both dimensions, the constants $c$ are real (complex) for the EW (MH) class. Therefore, in MH class the correlations shall decay in a modulated oscillatory way for large $s$. As long as we know, these results have never been confronted with numerical simulations to confirm their universality.

\begin{figure}[t]
	\includegraphics[width=4.25cm]{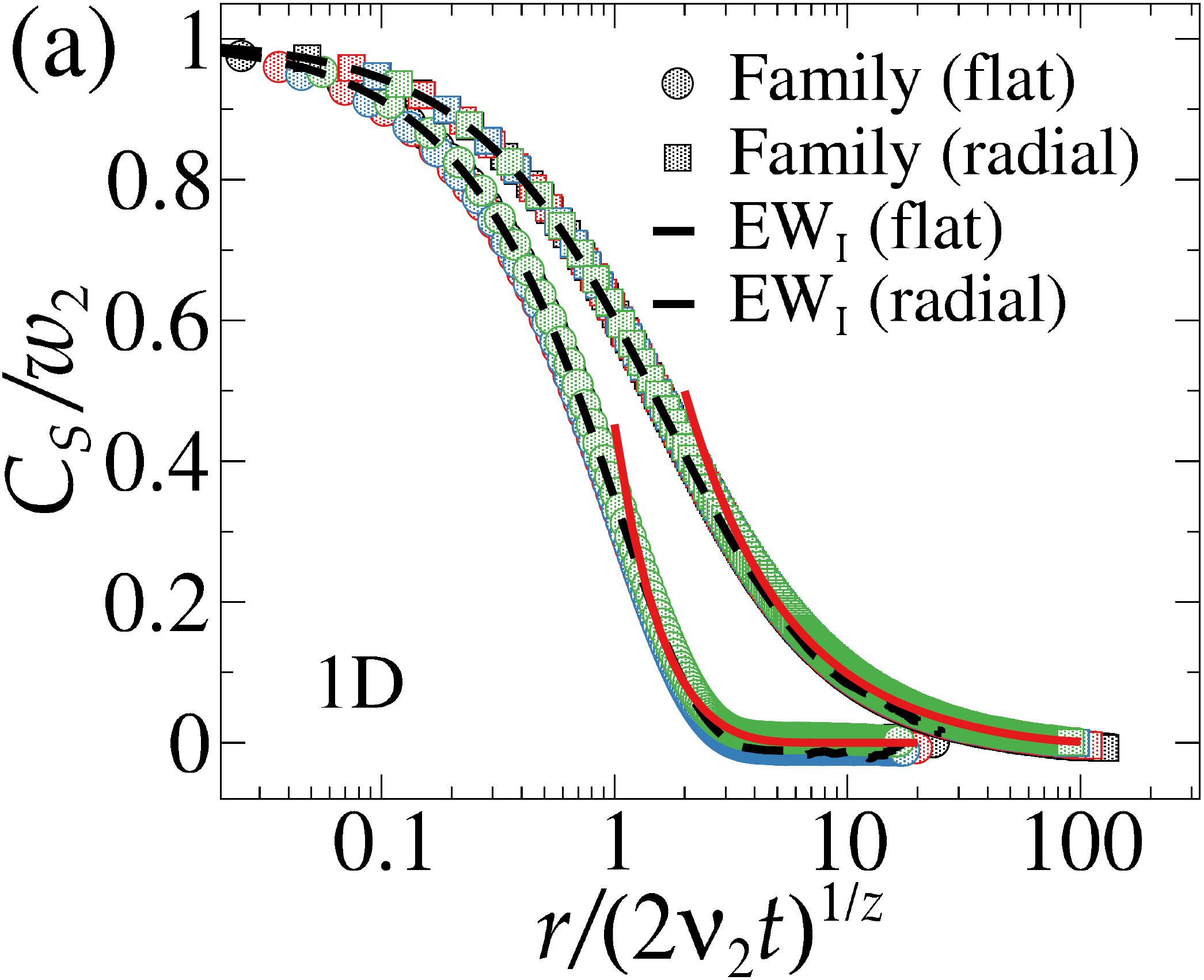}
	\includegraphics[width=4.25cm]{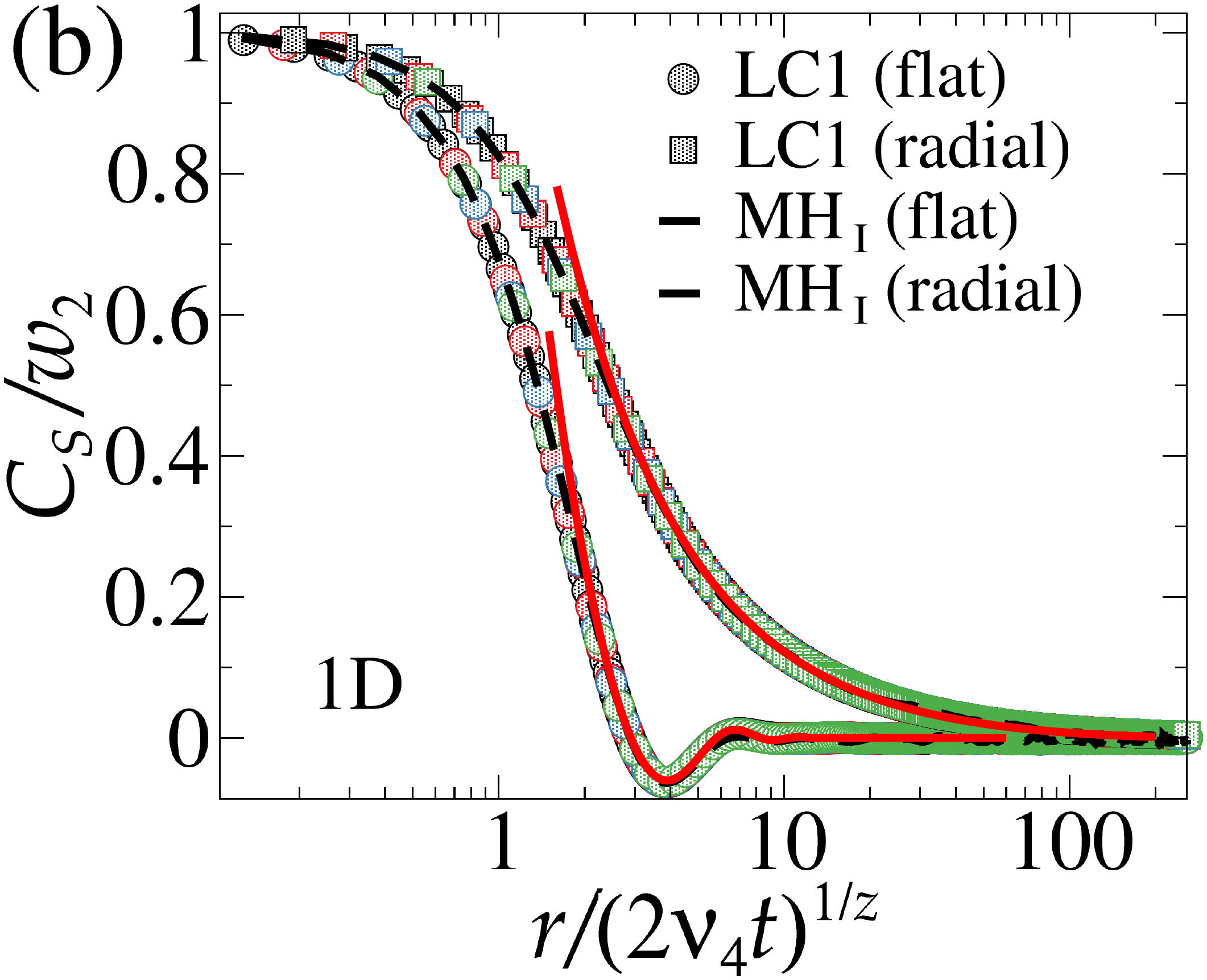}
	\includegraphics[width=4.25cm]{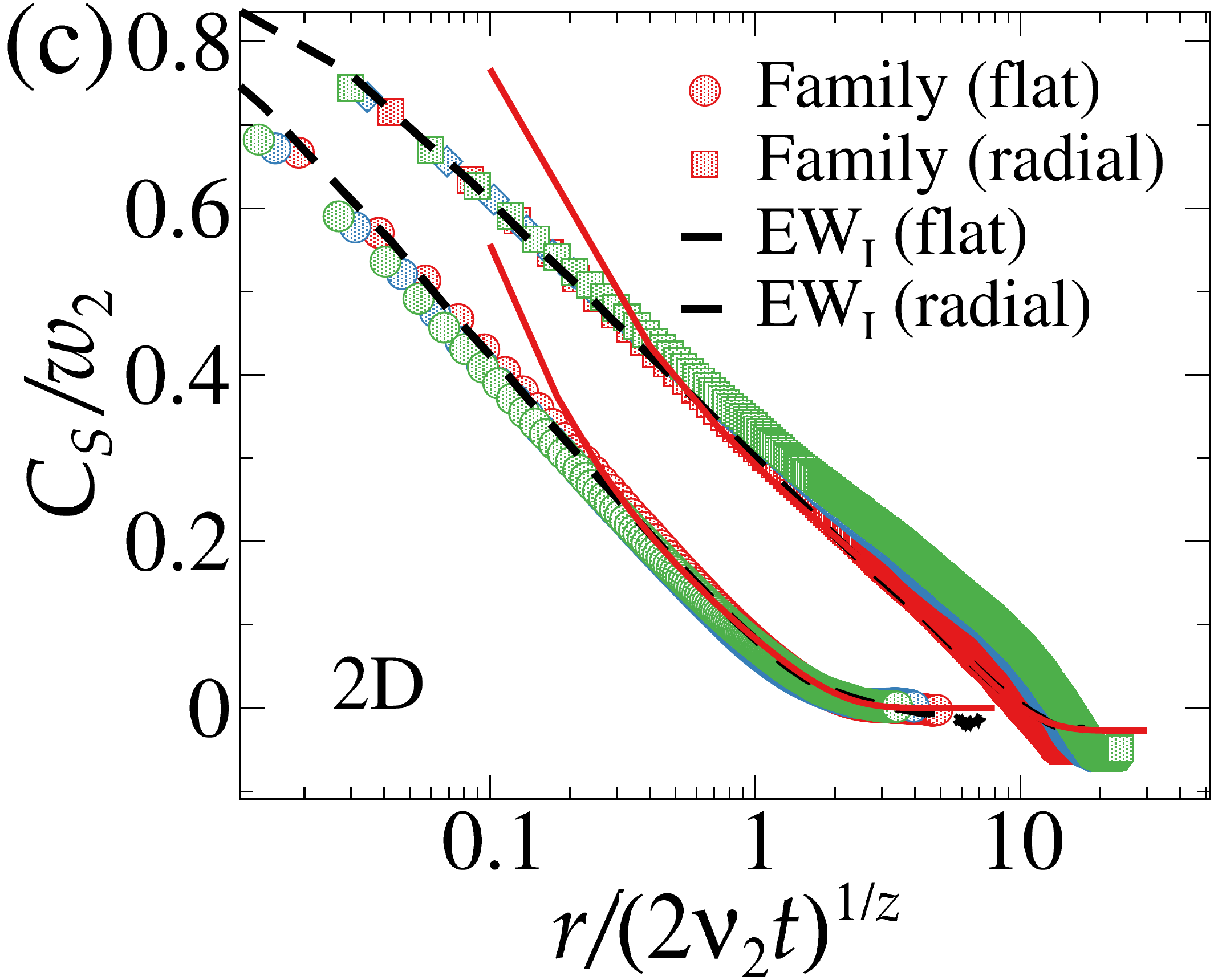}
	\includegraphics[width=4.25cm]{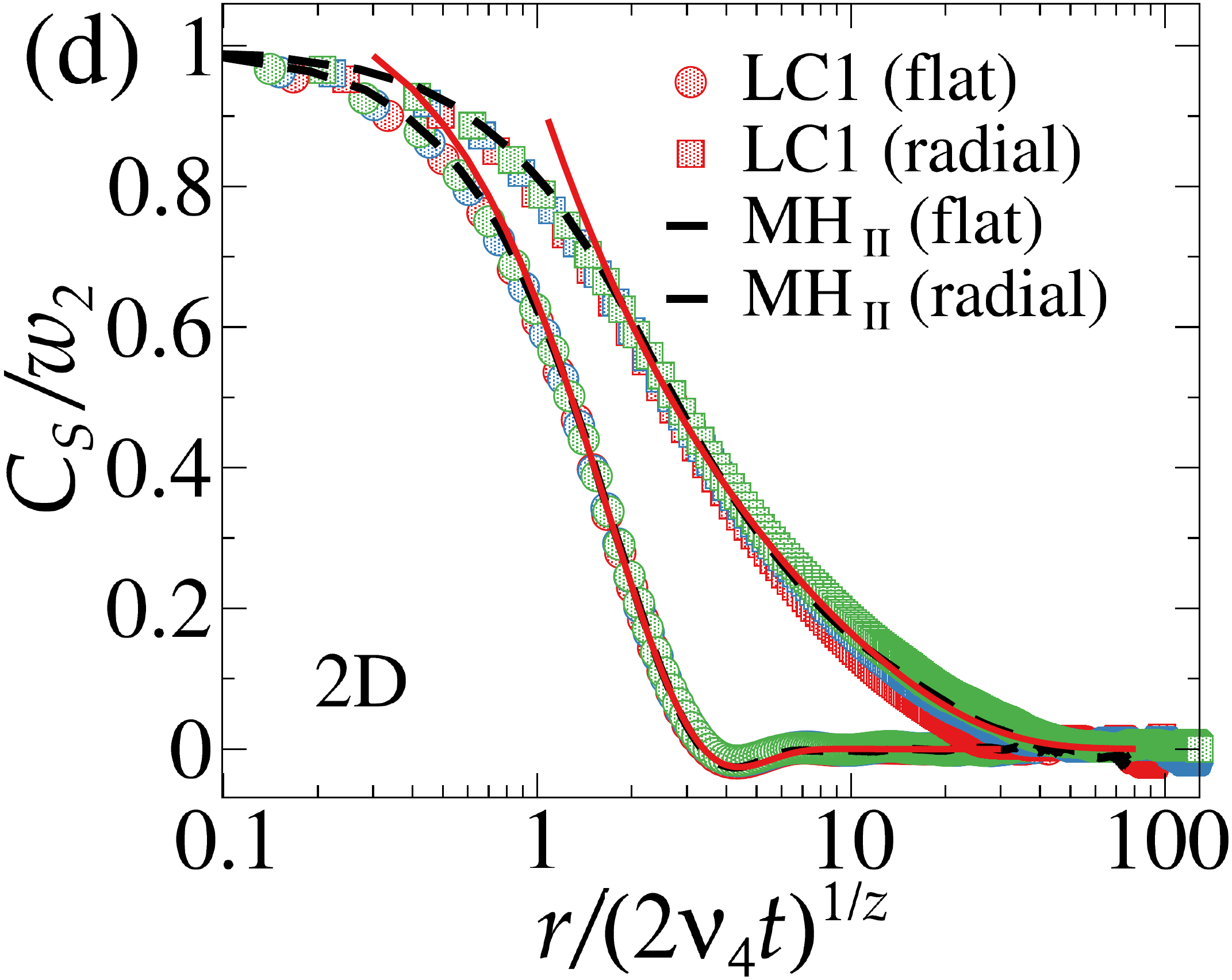}
	\caption{Rescaled spatial covariances for EW (left) and MH (right) classes, in 1D (top) and 2D (bottom). In all cases, results for the integrations of the respective growth equations are represented by dashed lines, while the symbols are results for the discrete models. In each panel, the top curves are data for expanding substrates and the bottom ones for fixed-size substrates. Different colors represent covariances measured at different times in the ranges $t \in [10^3,10^5]$ in 1D and $t\in [10^2,10^3]$ in 2D. The solid (red) lines are fits of the numerical data according to Eq. \ref{eqScalFuncC_S}.}
	\label{fig3}
\end{figure}

For the radial geometry in 1D, the correlation functions $C(\theta,\theta',t) \equiv \expct{\rho(\theta,t)\rho(\theta',t)}$ - where $\rho\equiv r-\expct{r}$ and $r(\theta,t)$ denotes the interface radius at polar coordinate $\theta$ - have been analytically investigated for the linear growth equations in \cite{Masoudi2}. However, only approximations for the case $\phi \equiv \theta-\theta' \ll 1$ were derived, which are not so useful here, once we want to compare the behavior of the entire covariance curves. Moreover, some expressions for the scaling function $F(s)$ were calculated for the 1D linear growth equations on growing domains \cite{Escudero2009}, but the interfaces analyzed there were not spatially homogeneous, due to the non-periodic boundary conditions considered, and, so, they are not expected to describe the circular interfaces we are interested in here. Some steps toward the calculation of the correlation functions for the spherical EW and MH equations were given in \cite{Escudero2009b}, but expressions for $F(s)$ have not been reported for this 2D radial geometry. Therefore, it seems that the effect of geometry on $F(s)$ for the linear UCs have never been demonstrated so far.

\begin{figure}[t]
	\includegraphics[width=4.25cm]{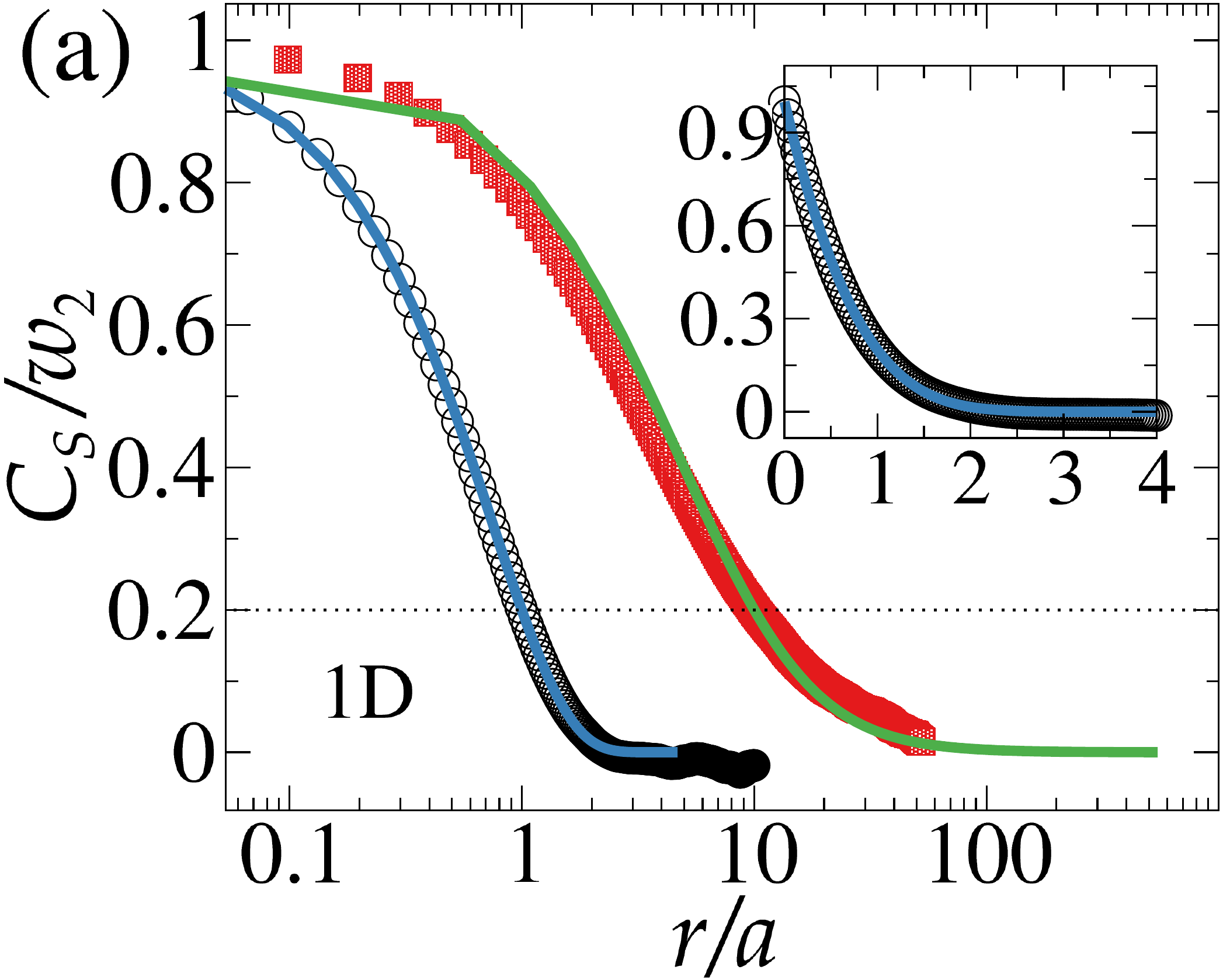}
	\includegraphics[width=4.25cm]{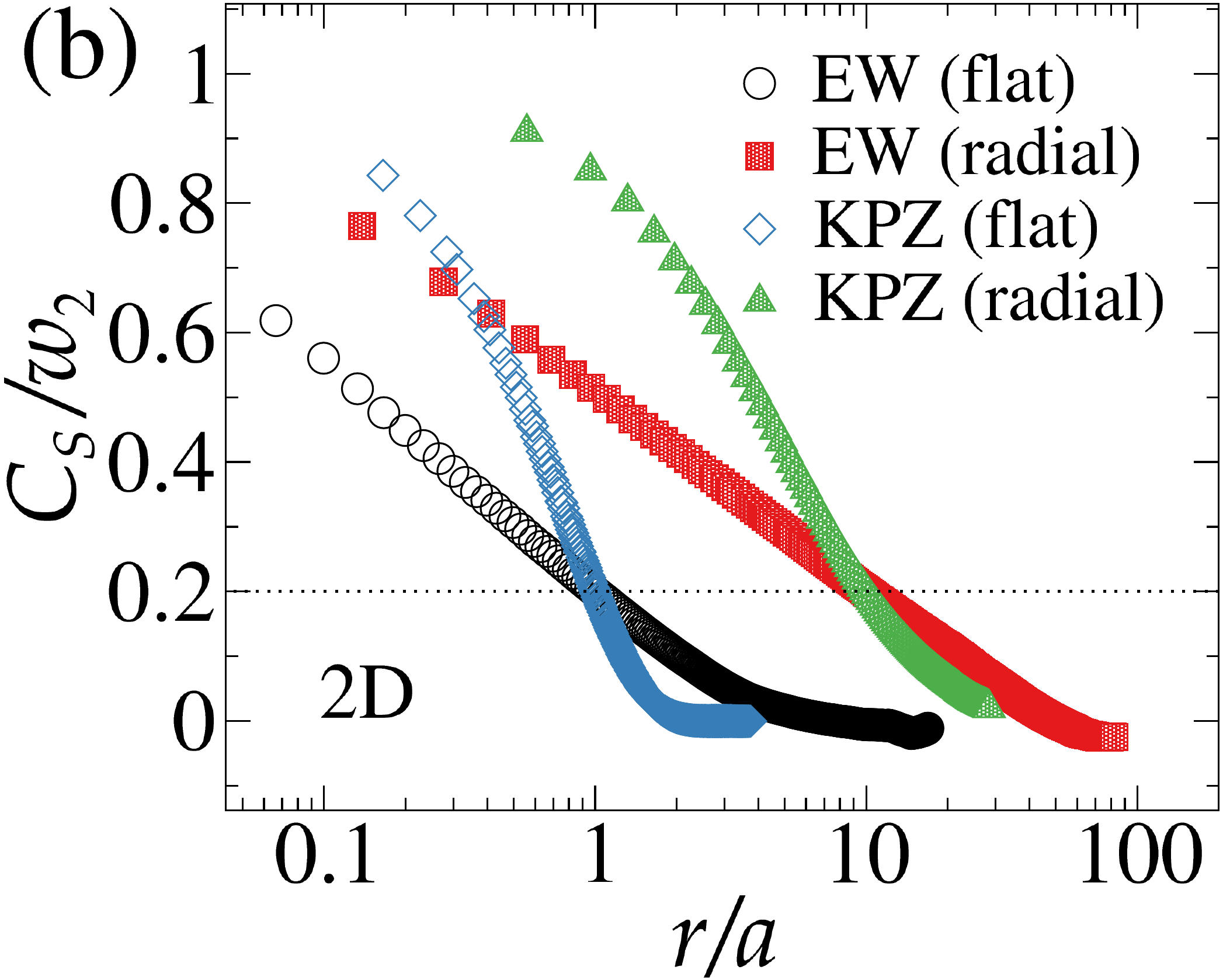}
	\includegraphics[width=4.25cm]{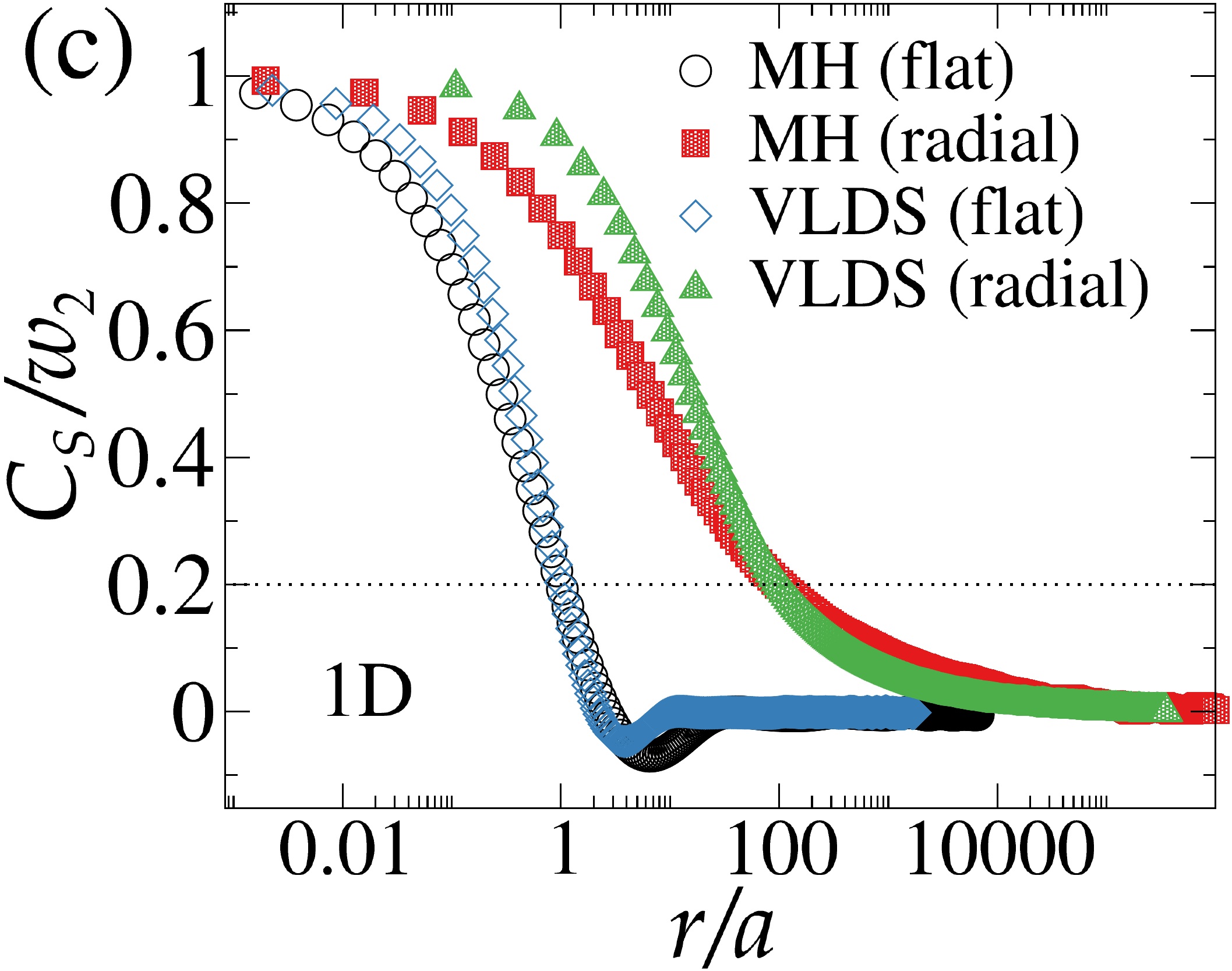}
	\includegraphics[width=4.25cm]{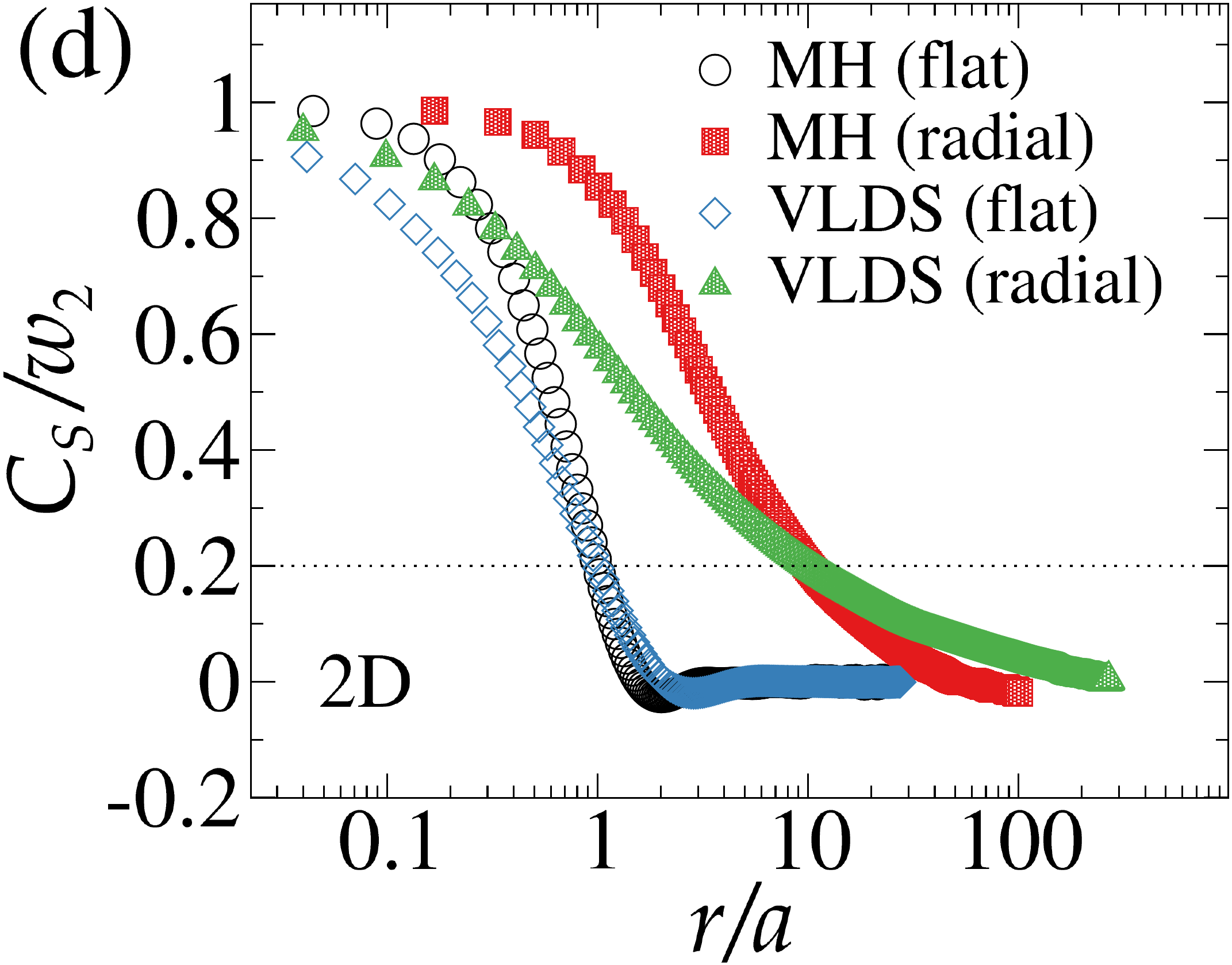}
	\caption{Comparison of rescaled spatial covariances for EW and KPZ [MH and VLDS] classes in 1D (a) [(c)] and 2D (b) [(d)]. In all cases, data for flat (open) and radial (full symbols) geometries are shown. In (a) the symbol and color schema is the same of (b); and the Airy$_1$ (solid) and Airy$_2$ (dashed) curves are the exact results for 1D KPZ class. The insertion presents the data for the flat EW class in 1D and the Airy$_1$ curve in linear scale. Results for the 2D KPZ class were extracted from \cite{Ismael14}, while the ones for the VLDS class came from \cite{Ismael16a}. In all panels, the scaling parameter $a$ was chosen to make the curves pass at $C_S/w_2 = 0.2$ (dotted lines) when $r/a=1$ for flat systems. For radial ones, the interceptions (at $C_S/w_2 = 0.2$) are placed at $r/a=10$, with exception of (c), where it is at $r/a = 100$.}
	\label{fig4}
\end{figure}

Figures \ref{fig3}(a)[(c)] and \ref{fig3}(b)[(d)] present the rescaled covariances respectively for EW and MH models in 1D [2D]. Firstly, we may note the striking collapse of curves for different models and deposition times for a given geometry, which gives strong evidence of their universality and also confirms the correctness of the coefficients $\nu_z$ estimated in the previous subsections for the discrete models. Secondly, it is clear that two different scaling functions $F(s)$ exist for fixed-size (flat) and expanding substrates (radial geometry). This is especially remarkable in the MH class, where the modulated oscillatory behaviors observed in the flat case gives place to simple monotonic decays in the radial geometry, in both 1D and 2D. We stress that this very same behavior was found by us for the VLDS class \cite{Ismael16a}, which is the nonlinear counterpart of the MH class. Moreover, while in EW class the decay is always monotonic, it is much slower in the radial case than in the flat one, a behavior which is also observed in the KPZ class (see e.g., \cite{Ismael14}) - the nonlinear counterpart of the EW class. 

These similarities lead us immediately to inquire whether the covariances for linear and nonlinear UCs might be the same, whose answer is almost always negative. Indeed, in Figs. \ref{fig4}(a)-(d) representative curves of the rescaled covariances for EW and MH classes in both dimensions and geometries are compared with the same data for their nonlinear counterparts (KPZ and VLDS, respectively). The results for 1D KPZ class are the exact Airy$_1$ and Airy$_2$ covariances for flat and radial systems, respectively \cite{Prahofer2002,Sasa2005,Borodin}. In 2D, the EW-KPZ and MH-VLDS curves are quite different. The MH-VLDS curves also present noticeable differences in 1D, as well as the 1D EW-KPZ covariances for the radial case. In the 1D flat case, however, the EW-KPZ curves present a remarkable collapse, as highlighted in the insert in Fig. \ref{fig4}(a). This strongly suggests that both have the same spatial covariance, namely, the EW covariance in 1D coincides with the Airy$_{1}$ one.

Let us now verify how Eq. \ref{eqScalFuncC_S} compares with our data. For the flat systems in 1D, we keep the exponents $\delta=z/(z-1)$ and $\gamma=1+\delta/2$ \cite{Majaniemi} fixed in Eq. \ref{eqScalFuncC_S} and try to fit the covariances using this equation with $c \in \mathbb{R}$ and the missing amplitude as free parameters in EW case. For MH class, since $c \in \mathbb{C}$ [i.e, $c = a + i b$, so that $F(s) \sim s^{-\gamma} e^{-a s^{\delta}} \cos\left( b s^{\delta} + \phi \right) $] more parameters are used to adjust the data. The resulting fittings are shown as red curves in Figs. \ref{fig3}(a) and \ref{fig3}(b), which are quite reasonable for $r/(2 \nu_z t)^{1/z} \gtrsim 1$. Interestingly, the covariances for both EW and MH classes with radial geometry in 1D are also well fitted by Eq. \ref{eqScalFuncC_S}, for large $r$, with $c \in \mathbb{R}$, $\delta=z/(z-1)$ and $\gamma=1$, as can be seen in Figs. \ref{fig3}(a) and \ref{fig3}(b). This confirms that the EW covariance for radial case is different from the Airy$_2$, since the last one decays asymptoticaly as an inverse square law \cite{Prahofer2002}. By assuming that the exponents $\delta$ are still the same in 2D (as suggested in \cite{Majaniemi}), we find a good agreement between the covariances for the integration of the 2D EW equation and Eq. \ref{eqScalFuncC_S} with $\gamma \approx 0.67$ and $\gamma \approx 0.39$ for the flat and radial geometry, respectively. See Fig. \ref{fig3}(c). For the integration of 2D MH equation, good fits are obtained in the flat case with $\gamma\approx 0$ and $c \in \mathbb{C}$, and in the radial case with $\gamma\approx 0.61$ and $c \in \mathbb{R}$ [see Fig. \ref{fig3}(d)].

\section{Temporal covariance}
\label{secCovT}

\begin{figure}[t]
	\includegraphics[width=4.25cm]{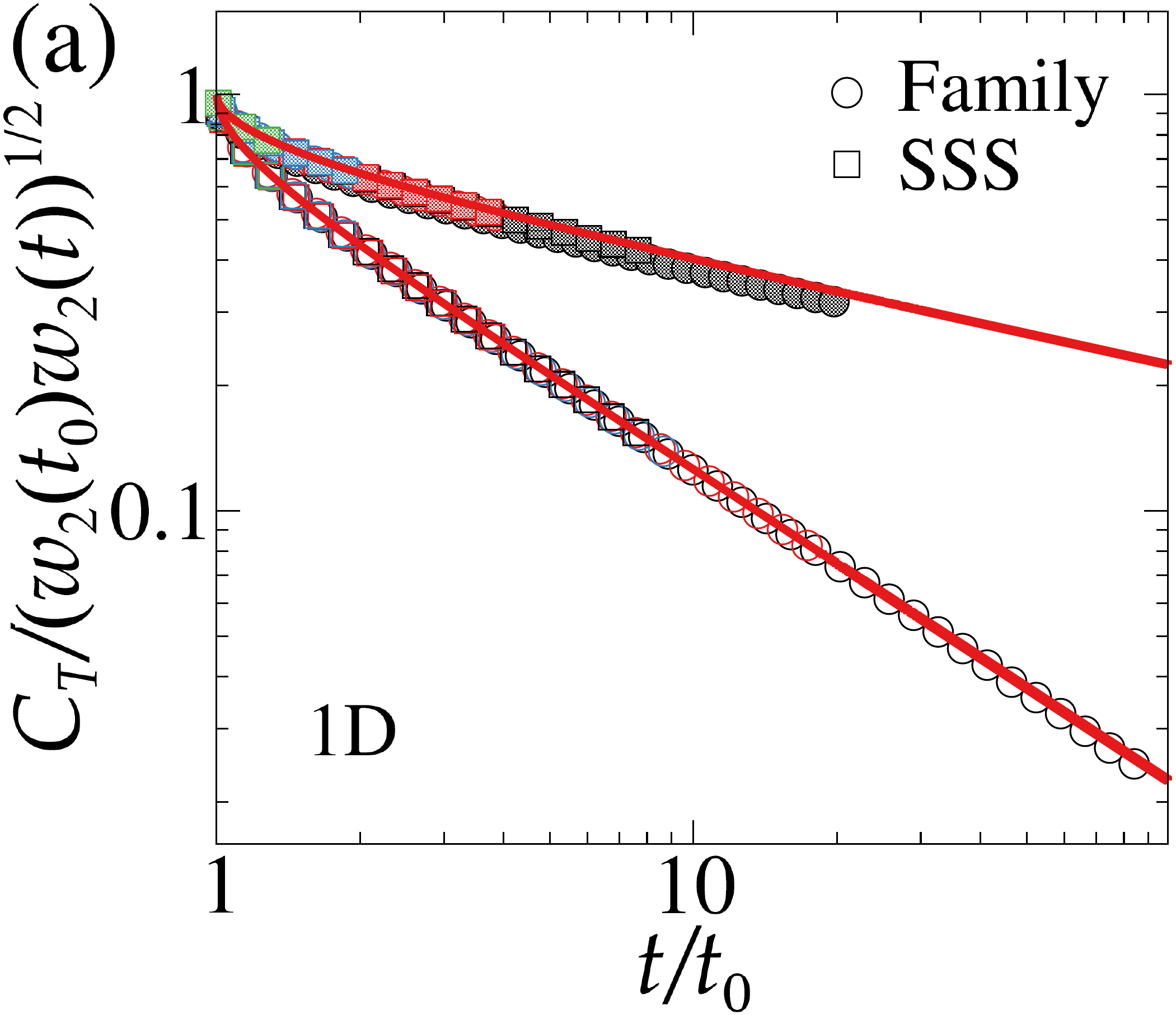}
	\includegraphics[width=4.25cm]{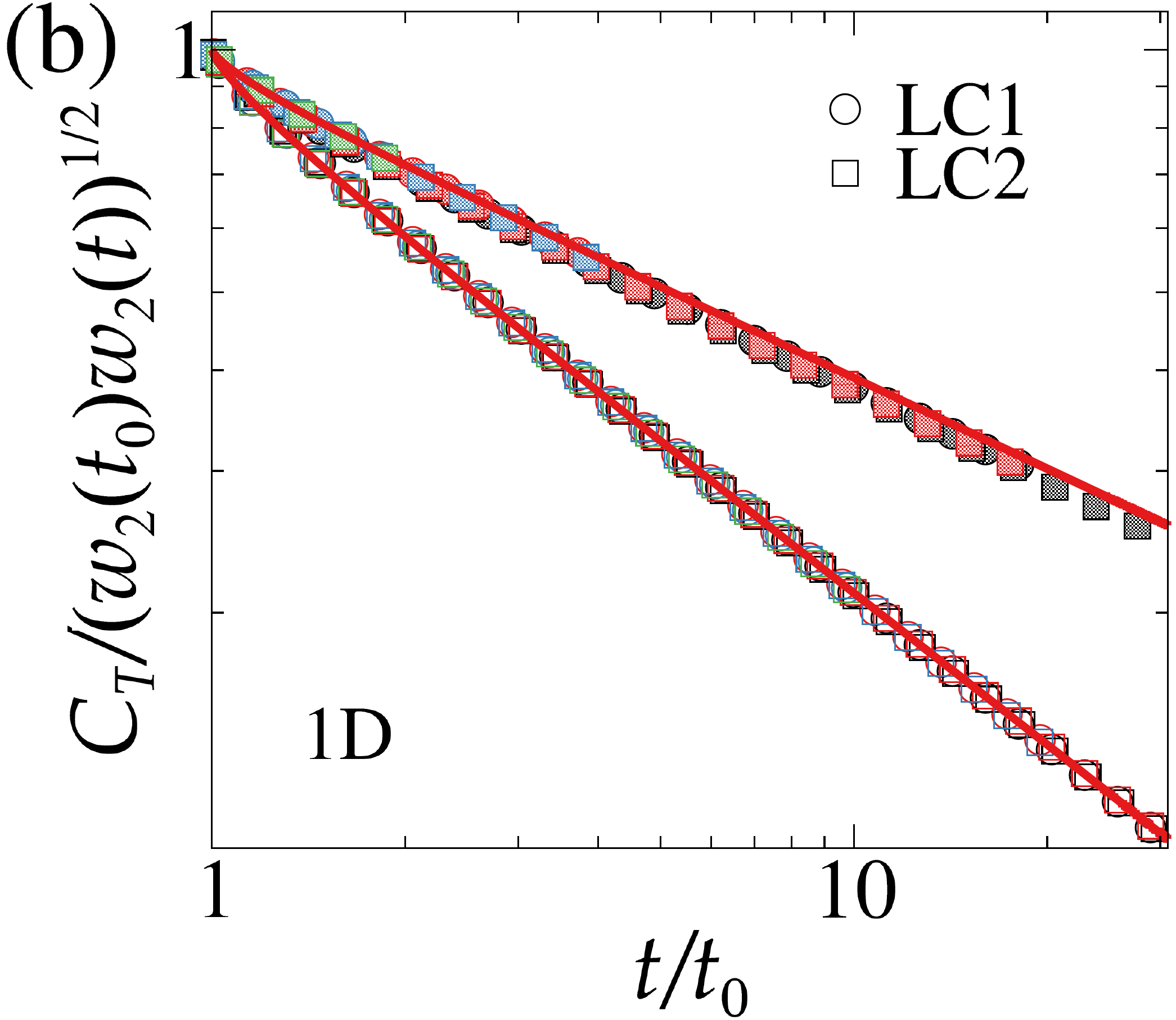}
	\includegraphics[width=4.25cm]{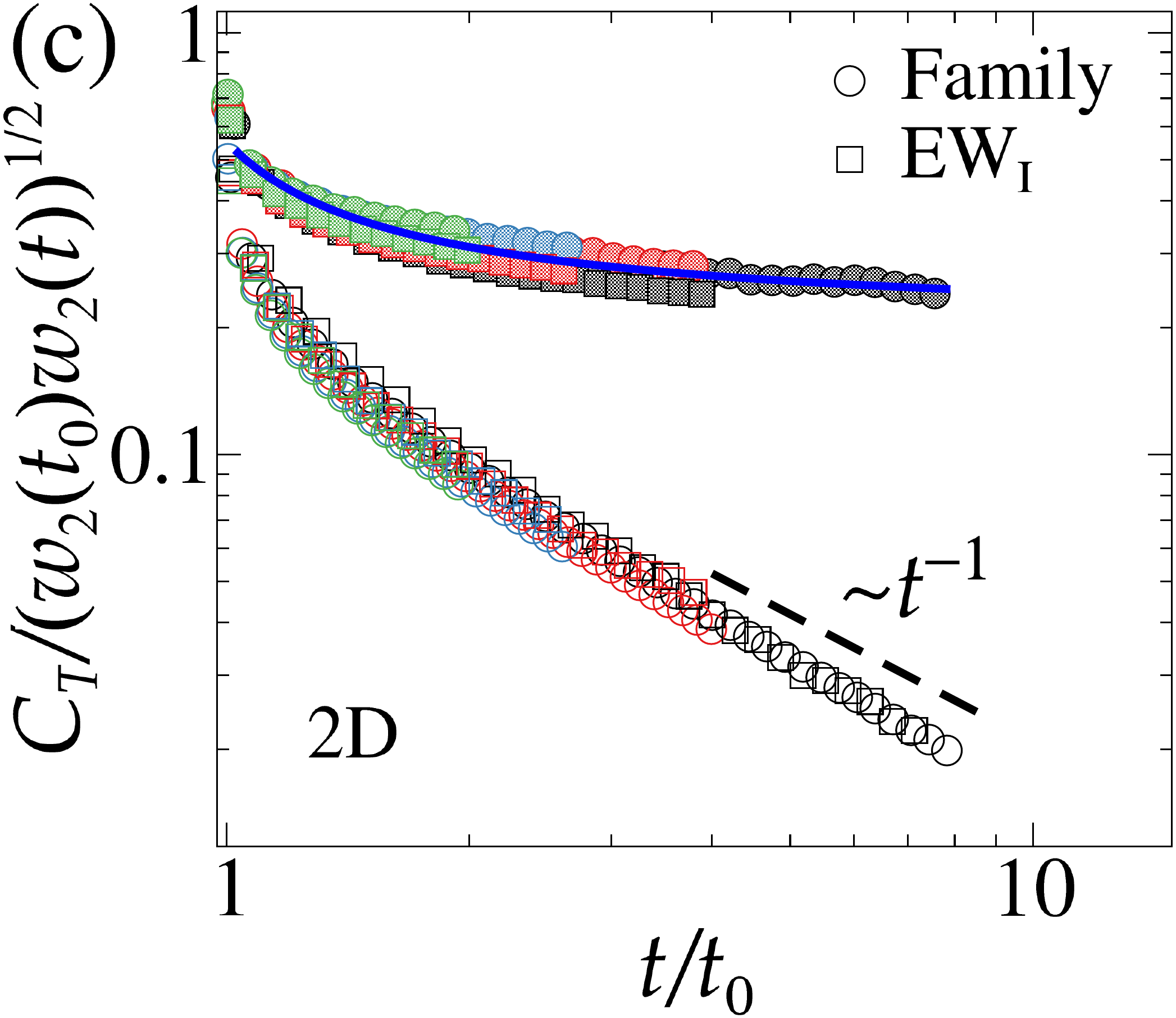}
	\includegraphics[width=4.25cm]{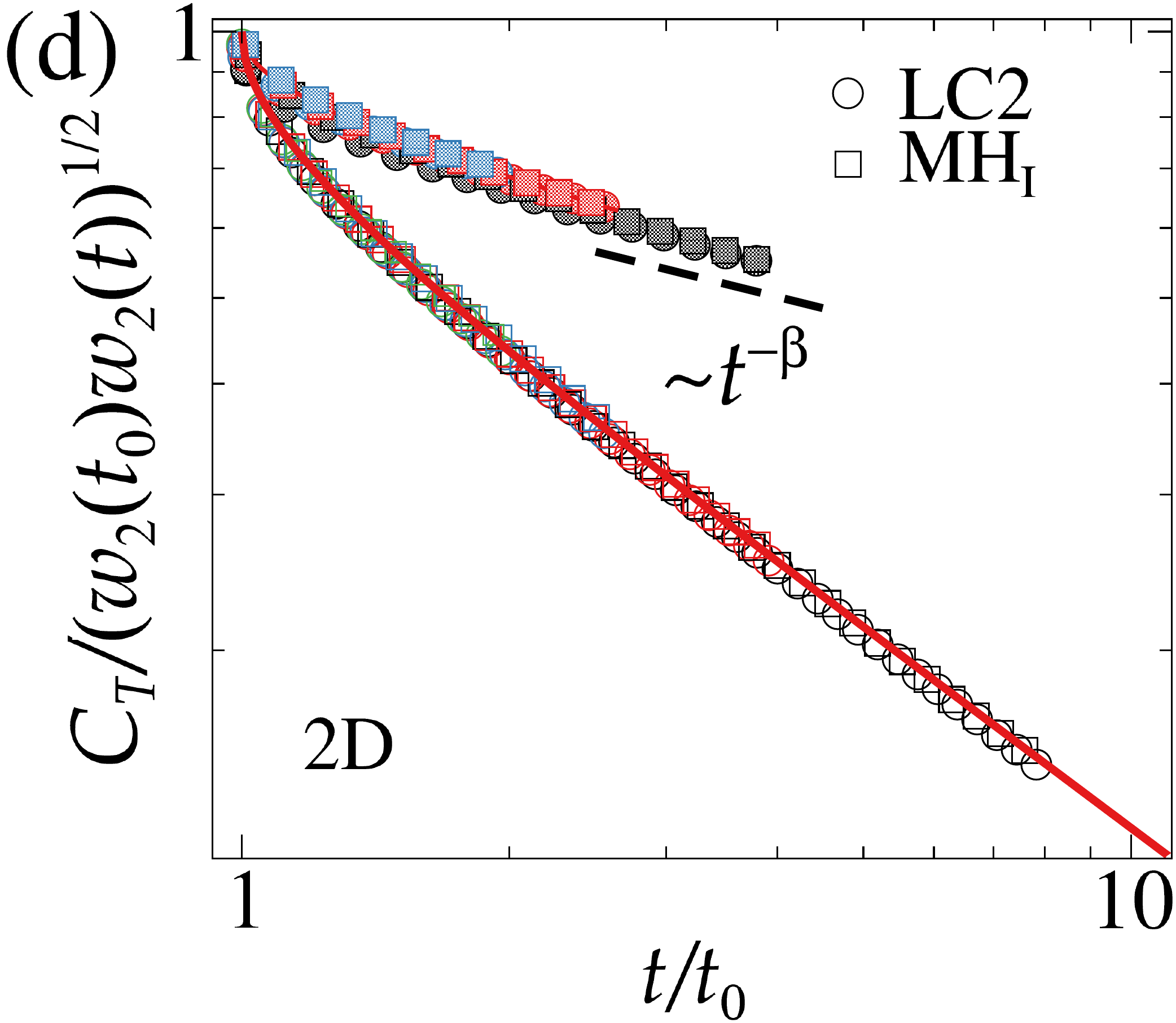}
	\caption{Rescaled temporal covariances for EW (left) and MH (right) classes, in 1D (top) and 2D (bottom). In all cases, the symbols are data for discrete models, being the open (closed) ones for fixed-size (expanding) substrates. Different colors represent different values for $t_0$, considered in the ranges $t_0 \in [10^3,10^4]$ in 1D and  $t_0 \in [10^2,10^3]$ in 2D. The continuous red lines are the exact analytical results, while the blue one in (c) is the fit discussed in the text. The dashed lines have the indicated slopes.}
	\label{fig5}
\end{figure}

Finally, we discuss the effect of geometry on the temporal covariance $C_T(t,t_0) = \left\langle \tilde{h}(x,t_0) \tilde{h}(x,t) \right\rangle \simeq \sqrt{w_2(t)w_2(t_0)} \mathcal{A}(t/t_0)$. As demonstrated by Krug \textit{et al.} \cite{KrugKallabis97}, for linear flat interfaces $C_T \sim [(t+t_0)^{2\beta}-(t-t_0)^{2\beta}]$, for $t > t_0 \gg 1$ for both 1D and 2D \cite{KrugKallabis97}, so that $\mathcal{A}^{f}(y) = (4 y)^{-\beta} [(y+1)^{2\beta} - (y-1)^{2\beta}]$. From solutions of the radial EW and MH equations in 1D by Singha \cite{Singha2005}, one knows also that $\mathcal{A}_{EW}^{r}(y) = \frac{2\sqrt{2}}{\pi} y^{-1/4} (1+1/y)^{-1/2} \sin^{-1}\sqrt{\frac{1+1/y}{2}}$ and $\mathcal{A}_{MH}^{r}(y) =  \frac{2\sqrt{2}}{\pi} y^{-3/8} {}_{2}F_{1}\left(\left[ \frac{1}{4},\frac{1}{4}\right],\left[ \frac{5}{4}\right];\frac{1+y^{-3}}{2} \right)$, where ${}_{2}F_{1}\left(\left[ a,b \right],\left[c\right];y\right)$ is the hypergeometric function. Curiously, it seems that no one of these results have never been compared with numerical data for discrete models. This is done here in Figs. \ref{fig5}(a) (for EW) and \ref{fig5}(b) (for MH class in 1D), where the collapse of the rescaled covariance curves for different models and $t_0$'s with the analytical expressions above let no room for doubt about their universality. The same is observed for the flat MH class in 2D [see Fig. \ref{fig5}(d)].

For the 2D radial geometry, as far as we know, there exist no analytical expressions for these covariances. Moreover, for the EW class in 2D, even in the flat case the covariance behavior is not known analytically. Nonetheless, similarly to what happens also in 1D, they are expected to decay asymptotically as $\mathcal{A} \sim (t_0/t)^{\bar{\lambda}}$ with $\bar{\lambda}=\beta+d_s/z$ in flat \cite{Kallabis99} and $\bar{\lambda}=\beta$ in radial systems \cite{Singha2005}. In fact, for large $t/t_0$ one observes exponents consistent with $\bar{\lambda} = 1$ for the flat EW class [see Fig. \ref{fig5}(c)] and $\bar{\lambda} = \beta$ for the radial MH class [Fig. \ref{fig5}(d)] in 2D. For the radial 2D EW class $\bar{\lambda}=\beta=0$, so that one could expect some kind of logarithmic decay of $\mathcal{A}(y)$. Unfortunately and certainly due to logarithmic corrections, in this case the collapse of data for different models and $t_0$'s is not so good as in the other systems. Anyhow, one finds that the average data is reasonable well fitted by $\mathcal{A}(y) = a + b/(c+\ln y)$, using $a$, $b$ and $c$ as fitting parameters. Despite this caveat, our results provide strong evidence that these temporal covariances are universal also in 2D.

\section{Conclusion}
\label{secConc}

In summary, we have demonstrated through extensive numerical analyses, as well as some analytical calculations that linear interfaces belonging to EW and MH classes split into subclasses depending on whether they are flat or radial. This happens even at the upper critical dimension, as is the case of EW in 2D. Although the 1-point fluctuations for all classes, subclasses and dimensionalities investigated are Gaussian, the universal and geometry-dependent character of the height distributions (HDs) manifests in their different variances $\expct{\chi^2}_c$. At this point, we remark that once the $\expct{\chi^2}_c$ values for EW and MH classes in a given geometry and dimension have considerable differences, they might serve as a useful indicator to decide between these classes in interfaces with Gaussian statistics. By the same token, the remarkable differences in the spatial and temporal covariances and their universality (as demonstrated here for the linear cases and elsewhere for the nonlinear ones) confirm that these are very important quantities to determine the class of a given evolving interface. An interesting exception to this is the spatial covariance for the flat EW class in 1D, which display an unanticipated agreement with the rescaled curve for the flat 1D KPZ class, strongly indicating that 1D flat EW interfaces are also described by the so-called Airy$_1$ process. This result, which is likely related to the fact that EW and KPZ flat interfaces share the same statistics at the steady state regime in 1D, shall motivate theoretical studies to verify the relevance of the KPZ nonlinearity in such quantity. Anyhow, in the 2D case, which is the most important for applications, the spatial covariances are always quite different. For instance, HDs and spatial covariances have been employed in recent works to confirm the universality class in thin film growth by vapor deposition \cite{Almeida14,healy14exp,Almeida15,Almeida17} and electrodeposition \cite{Yuri15,Rodolfo17}, despite the lacking of a confirmation of their universality in the linear case so far. Therefore, our results are appealing from this applied point of view. Furthermore, together with all existing studies for the nonlinear case (already quoted in the Introduction), the present work seals the deal about geometry-dependence in interface growth for the most important (KPZ, VLDS, EW and MH) classes.

\acknowledgments

This work was supported by CNPq, CAPES, FAPEMIG and FAPERJ (Brazilian agencies). We thank F. Bornemann for kindly providing the covariances of the
Airy processes.

\appendix

\section{Derivation of the variance of the HDs in radial case for general $z$}
\label{apExSol}

The linear growth equation in polar coordinates, for general even dynamic exponent $z$, in 1D, is given by \cite{Singha2005}
\begin{equation}
 \frac{\partial r(\theta,t)}{\partial t} = F - (-1)^{z/2} \frac{\nu}{r^z} \frac{\partial^z r(\theta,t)}{\partial \theta^z} + \eta(\theta,t),
 \label{eqLinRadial}
\end{equation}
where $\expct{\eta(x,t)} = 0$ and $\expct{\eta(\theta,t)\eta(\theta',t')} = (2D/r) \delta(\theta-\theta')\delta(t-t')$. It can be linearized by considering that fluctuations are small, so that $r^z$ can be replaced by $\expct{r}^z$, with  $\expct{r} = r_0 + F t$, where $r_0$ is the initial radius. This yields
\begin{equation}
 \frac{\partial r(\theta,t)}{\partial t} = F - (-1)^{z/2} \frac{\nu}{(r_0 + Ft)^z} \frac{\partial^z r(\theta,t)}{\partial \theta^z} + \eta(\theta,t).
\end{equation}
The same approximation can be done in the noise correlation: $\expct{\eta(\theta,t)\eta(\theta',t')} = [2D/(r_0+Ft)] \delta(\theta-\theta')\delta(t-t')$. Now, if $\tilde{r}_n(t)$ and $\tilde{\eta}_n(t)$ denote the Fourier transforms of $r(\theta,t)$ and $\eta(\theta,t)$, respectively, we arrive at
\begin{equation}
 \frac{\partial \tilde{r}_n(t)}{\partial t} = - \frac{\nu n^z}{(r_0 + Ft)^z} \tilde{r}_n(t) + \tilde{\eta}_n(t),
\end{equation}
from which the solution for $\delta \tilde{r}_n = \tilde{r}_n - \expct{\tilde{r}_n}$ is obtained, being
\begin{multline}
 \delta \tilde{r}_n(t) = e^{\nu n^z/[(z-1)F(r_0+Ft)^{z-1}]} \\\times\int_0^{t} dt' e^{-\nu n^z/[(z-1)F(r_0+Ft')^{z-1}]}.
\end{multline}
The squared interface width is given by
\begin{equation}
 w_2(t) = \frac{1}{2\pi} \int_0^{2\pi} \expct{\delta r(\theta,t)} d\theta = \sum_{n=-\infty}^{\infty} \expct{\delta \tilde{r}_n(t)\delta \tilde{r}_{-n}(t)},
\end{equation}
which yields
\begin{eqnarray}
 w_2 = \frac{D}{\pi} \sum_{n=-\infty}^{\infty} && e^{2\nu n^z/[(z-1)F(r_0+Ft)^{z-1}]} \\ \nonumber
 &&\times \int_0^{t} dt' \frac{e^{-2\nu n^z/[(z-1)F(r_0+Ft')^{z-1}]}}{r_0 + Ft'}.
\end{eqnarray}
Now, assuming that $r_0=0$, we find
\begin{eqnarray}
 w_2 = \frac{D}{F\pi} \int_0^{t} \frac{dt'}{t'} \sum_{n=-\infty}^{\infty} e^{-\left( \frac{n}{\sigma(t')}\right)^z },
 \label{eqAprug}
\end{eqnarray}
where 
\begin{equation}
\sigma(t') \equiv F t^{\frac{z-1}{z}} \left(\frac{z-1}{2\nu}\right)^{1/z} \frac{t'^{\frac{z-1}{z}}}{(t^{z-1}-t'^{z-1})^{1/z}}.
\end{equation}
Using the Poisson summation formula
\begin{multline}
S_z \equiv \sum_{n=-\infty}^{\infty} e^{-\left( \frac{n}{\sigma}\right)^z } = \sum_{k=-\infty}^{\infty} \int_{-\infty}^{\infty} dx e^{-\left( \frac{x}{\sigma}\right)^z -2 \pi i k x},
\end{multline}
where the integral can be exactly calculated for a given $z$ with the help of an algebra software and it is related to summations of hypergeometric functions. Assuming that the main contribution for this sum come from $k=0$, one finds $S_z = c_z \sigma$, where $c_z$ is a constant (for instance, $c_2=\sqrt{\pi}$, $c_4 = \frac{\pi}{\sqrt{2} \Gamma(3/4)}$, $c_6 = \frac{2 \pi}{3 \Gamma(5/6)}$ and so on). Then, returning to Eq. \ref{eqAprug} we obtain
\begin{equation}
 w_2(t) = \left( \frac{D}{\nu^{1/z}} \right) \left[ \frac{c_z (z-1)^{1/z} I_z}{2^{1/z} \pi} \right] t^{\frac{z-1}{z}},
\end{equation}
where
\begin{equation}
I_z \equiv \int_0^t dt' \frac{t'^{-1/z}}{(t^{z-1}-t'^{z-1})^{1/z}},
\end{equation}
which can be again easily calculated, being $I_2=\pi$, $I_4=\sqrt{2}\pi/3$, $I_6=2\pi/5$ and so on. Finally, bearing in mind the ``KPZ ansatz'' (Eq. \ref{Eqansatz}), we have that $w_2 = \Theta^{2\beta} \expct{\chi^2}_c t^{2\beta}$, where $\beta = (z-1)/2 z$ is indeed the expected growth exponent \cite{barabasi,KrugAdv}. Moreover, we may identify $\Theta^{2\beta} = D/\nu^{1/z}$, meaning that $\Theta = (D^{z}/\nu)^{\frac{1}{z-1}}$ and, thus, we have the variance of the HDs
\begin{equation}
 \expct{\chi^2}_c = \frac{c_z (z-1)^{1/z} I_z}{2^{1/z} \pi}.
\end{equation}
For the EW class ($z=2$) this gives $\expct{\chi^2}_c = \sqrt{\pi/2} \approx 1.25331$, as already reported in \cite{Singha2005,Masoudi1}. For the MH class ($z=4$), which is our main interest in this calculation, one finds $\expct{\chi^2}_c = \pi/[54^{1/4}\Gamma(3/4)] \approx 0.94573$.

\section{Calculation of the phenomenological coefficients with the inverse method}
\label{apInvMet}

Following \cite{Lam}, the inverse method consists in taking a giving (initial) interface at time $t$ and, starting from it, evolving $m$ different realizations of the growth during a time interval $\Delta t$. The average of these $m$ interfaces generates a height profile which minimizes the local effects of the noise, while it keeps (and uncovers) the main features of the long wavelength fluctuations. Therefore, the parameters of a candidate equation (e.g, $F$, $\nu_z$ and $D$ in Eqs. \ref{eqEW} and \ref{eqMH}) can be obtained by minimizing the difference between the evolution of the initial profile predicted by the given equation and the average profile measured. Considering a time interval $\Delta t$, the evolution can be approximated by
\begin{equation}
\frac{\langle\Delta h(x,t)\rangle}{\Delta t}\approx \vec{a}\cdot\vec{H},
\end{equation}
where the vector $\vec{a}$ contains the set of phenomenological parameters of the equation and $\vec{H}$ is the deterministic derivatives related to the relaxation process. The parameters are obtained minimizing the cost function
\begin{equation}\label{costfunc}
S=\frac{1}{N}\sum^N_{i=1}\left(\frac{\langle \Delta h_i \rangle}{\Delta t}-\vec{a}\cdot\vec{H}\right)^2.
\end{equation}

Usually, this minimization would lead to a matrix equation. However, since in our case the only unknown coefficient is $\nu_z$, the minimization of the previous equation leads to
\begin{equation}\label{inverso_direto}
\nu_z=\frac{\sum^N_{i=1}(\langle \Delta h_i\rangle/\Delta t-1)\partial^z_x h}{\sum^N_{i=1}(\partial^z_ x h)^2},
\end{equation}
with $z=2$ for EW and $z=4$ for MH class.

\begin{figure}[!t]
	\centering
	\includegraphics[width=7.5cm]{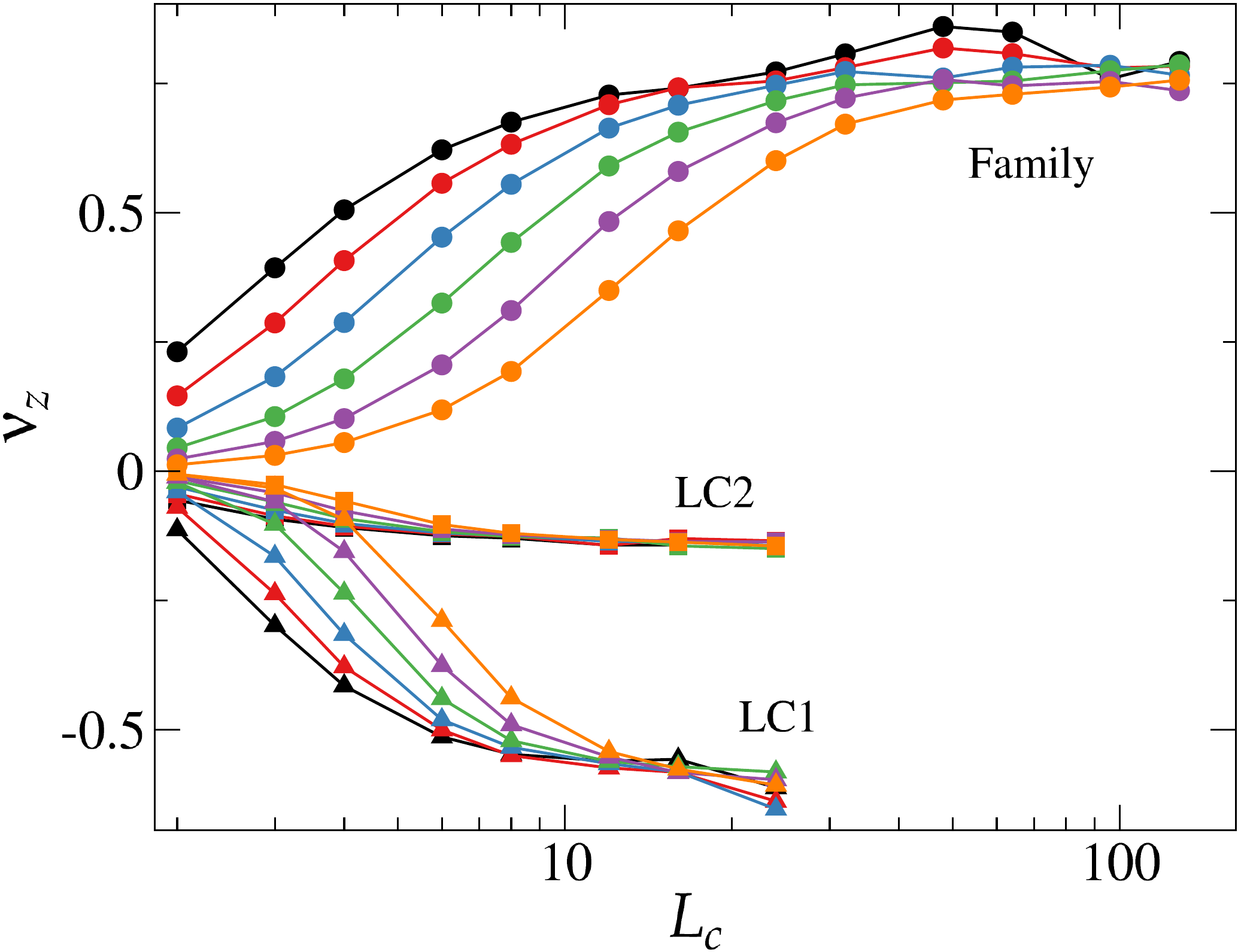}
	\caption{Calculation of the $\nu_z$ for the different models in 1D through the inverse method. The different curves represent data for $\Delta t=2$ (black), $4$ (red), $8$ (blue), $16$ (green), $32$ (purple) and $64$ (orange). The lines are guide to the eye.}
	\label{fig6}
\end{figure}

Finally, since the function $h$ presents spatial variations in the scale of its discretization, the application of discrete differentiations would lead to imprecise results. To avoid this problem, the derivatives are applied in the Fourier space after some coarse-graining, which is performed by truncating wavelengths smaller than $L_c$. Note that, even though the parameters in the linear classes does not change under renormalization, in this discrete environment the result of differentiations changes over the coarse-graining scale.

The parameters obtained through this slight variation of the inverse method can be seen in Fig. \ref{fig6}. It shows the values of $\nu_z$ as a function of $L_c$ for different $\Delta t$'s. For large enough $L_c$, the values obtained by different $\Delta t$'s agree, from which we estimate the values of $\nu_z$ displayed in Tab. \ref{tab1}. It is important to notice that the faster is the increasing of the interface width, the better are the estimates coming from the inverse method. Since in 2D such increase is very slow for the systems investigated here, the application of the inverse method is not worthy. 


\bibliography{biblinearclass_split}

\end{document}